\documentclass[10pt,a4paper]{article}

\usepackage{amsmath,amsfonts,latexsym,amssymb,amsthm}
\usepackage{verbatim,amsthm,curves,graphics}
\usepackage{mathrsfs}

\begin{document}

\theoremstyle{plain}
\newtheorem{thm}{Theorem}[section]
\newtheorem{lem}{Lemma}[section]
\newtheorem{prop}{Proposition}[section]

\theoremstyle{definition}
\newtheorem{defn}{Definition}[section]
\newtheorem{remark}{Remark}[section]

\renewcommand{\thesection}{\Roman{section}}
\newcommand{\esssup}{\operatorname*{ess.sup}}
\newcommand{\mr}{\mathbb{R}} 
\newcommand{\mz}{\mathbb{Z}} 
\newcommand{\mh}{\mathbb{H}} 
\newcommand{\ms}{\mathbb{S}} 
\newcommand{\mc}{\mathbb{C}} 
\newcommand{\mn}{\mathbb{N}} 
\newcommand{\rSpin}{\operatorname{Spin}}  
\newcommand{\rSO}{\operatorname{SO}}      
\newcommand{\rO}{\operatorname{O}}        
\newcommand{\rCliff}{\operatorname{Cliff}}
\newcommand{\WF}{\operatorname{WF}}       
\newcommand{\Pol}{{\rm WF}_{pol}}         
\newcommand{\cO}{\mathscr{O}}             
\newcommand{\cA}{\mathscr{A}}             
\newcommand{\cR}{\mathscr{R}}             
\newcommand{\cC}{\mathscr{C}}             
\newcommand{\cF}{\mathscr{F}}             
\newcommand{\cK}{\mathscr{K}}             
\newcommand{\cH}{\mathscr{H}}
\newcommand{\cN}{\mathscr{N}}
\newcommand{\cI}{\mathscr{I}}
\newcommand{\cQ}{{\cal Q}}
\newcommand{\cD}{\mathscr{D}}             
\newcommand{\cL}{\mathscr{L}}             
\newcommand{\bS}{\operatorname{S}}
\newcommand{\pgator}{/\!\!\!S}
\newcommand{\car}{\operatorname{CAR}}
\newcommand{\bL}{\operatorname{OP}}
\newcommand{\bU}{\mathcal{U}}
\newcommand{\singsupp}{\operatorname{singsupp}}
\newcommand{\tr}{\operatorname{tr}}
\newcommand{\rd}{{\rm d}}                 
\newcommand{\mslash}{/\!\!\!}             
\newcommand{\slom}{/\!\!\!G}         
\newcommand{\dirac}{/\!\!\!\nabla}        
\newcommand{\myid}{\leavevmode\hbox{\rm\small1\kern-3.8pt\normalsize1}}
\renewcommand{\Im}{\operatorname{Im}}
\renewcommand{\Re}{\operatorname{Re}}

\title{The Hadamard Condition for Dirac Fields and Adiabatic States on
       Robertson-Walker Spacetimes}
\author{Stefan Hollands\footnote{Electronic mail:
        \tt sh128@york.ac.uk}\hspace{0.5em}\\
       \it{Department of Mathematics, University of York,} \\ 
       \it{York YO10~5DD, UK}}
\date{2 July 2001}

\maketitle

\begin{abstract}
We characterise the  homogeneous and isotropic gauge invariant and
quasifree states for free Dirac quantum fields on Robertson-Walker 
spacetimes in any even dimension. Using this characterisation, we
construct adiabatic vacuum states of order $n$ corresponding to 
some Cauchy surface. We then show that any two such states (of 
sufficiently high order) are locally quasi-equivalent. 
We propose a microlocal version of the Hadamard condition for spinor 
fields on arbitrary spacetimes, which is shown to entail the usual 
short distance behaviour of the twopoint function. The polarisation 
set of these twopoint functions is determined from the Dencker connection
of the spinorial Klein-Gordon operator which we show to equal the
(pull-back) of the spin connection. Finally it is demonstrated that 
adiabatic states of infinite order are Hadamard, and that those of order
$n$ correspond, in some sense, to a truncated Hadamard series and will 
therefore allow for a point splitting renormalisation of the 
expected stress-energy tensor.
\end{abstract}

\section{Introduction}

In many cases of physical interest, for example the early stages
of the universe or stellar collapse, one is naturally led to the problem of
constructing quantum field theories on a non-static curved spacetime.
Numerous papers have been devoted to the study of linear scalar fields
on such backgrounds, but less has been done for fields 
with higher spin, mainly because the analysis of multicomponent fields is 
technically more involved. The aim of the present paper is to partly 
fill this gap for the case of a Dirac field on 
a curved spacetime. 

Quantum field theory in curved spacetime (in short, QFT in CST) is best described
within the algebraic approach to quantum field theory, which started with the work
of Haag and Kastler \cite{haka}, for an overview see \cite{haag}. 
In this approach one deals with a net of $C^*$-algebras
$\{\cA(\cO)\}_{\cO \subset M}$ of observables localised in a 
spacetime region $\cO \subset M$. The algebra $\cA = 
\overline{\cup_{\cO \subset M} \cA(\cO)}$ is called the `quasilocal
algebra'. Quantum states in the algebraic framework are positive
normalised linear functionals on the quasilocal algebra. 
One of the major difficulties
of QFT on CST is to pick out physically reasonable states. 

It has become widely accepted by now that, for a linear scalar field,  
the so-called ``Hadamard states'' are good candidates for physical states. 
These states are distinguished among other states by the 
particular form of the singular part of their  
twopoint function. A mathematically precise definition 
was given in~\cite{kaywald}. 
The following facts about Hadamard states are known: 

They allow for a point-splitting renormalisation 
of the stress-energy tensor $T_{\mu\nu}$~\cite{wald}. 
Verch \cite{Verch} has shown that Hadamard states are 
locally quasi-equivalent and he has also 
shown local definiteness in the sense of Haag et. al. \cite{HNS}. 
Radzikowski \cite{Radzikowski} discovered that (quasifree) Hadamard
states, initially defined by the singular behaviour of the twopoint
function in position space, can also be characterised by the 
so-called `wave front set' of that twopoint function, a central concept 
in the mathematical subject called `microlocal analysis', 
aimed at describing the singular behaviour of distributions. 
(Since these techniques do not belong to the daily used toolchest of 
the theoretical physicist so far, we give a brief introduction to 
this subject in the appendix.) 
The microlocal characterisation of Hadamard states is much easier to check 
in many cases where an explicit expression of
the twopoint function cannot be obtained and has already led 
to important progress in the subject. It played an important r\^ole 
in the proof~\cite{RadzikowskiII} of Kay's conjecture~\cite{gk} in (axiomatic) QFT in CST, 
in the derivation of ``quantum inequalities''~\cite{fewster}
and for the perturbative construction of self-interacting quantum field
theories in general globally hyperbolic curved spacetimes~\cite{bf}.

\medskip

There is a pre-existing notion of a Hadamard state for Dirac fields in a curved
spacetime~\cite{koeh, ver}, analogous to the  
condition on the singular part of the twopoint function for a linear
scalar field. It had been expected that there should also be a microlocal 
characterisation as for the scalar field, 
but the details had never been spelled out.  One purpose of this paper
is to close this gap. The microlocal condition that we propose is
similar to that in the spin-0 case. But it differs in that, unless 
the state in question is assumed to be charge invariant, it needs to be 
imposed separately for the positive and negative frequency twopoint 
functions. 
We show (Thm.~\ref{hadachar}) that our microlocal notion of Hadamard states
coincides with the concept based on the short-distance behaviour
put forward in \cite{koeh, ver}. Our result is the counterpart of similar 
theorem by Radzikowski~\cite{Radzikowski} obtained earlier for a scalar field. 

New questions also arise in the spin-1/2 case that have no 
counterpart in the spin-0 case. For example, it is natural to ask what the 
most singular components of the twopoint function are. It appears that 
the natural mathematical setting to analyse this question is 
provided by the concept of the `polarisation set'~\cite{Denck} of 
a vector valued distribution (such as the twopoint function in 
the spin-1/2 case), a notion which refines that of the 
wave front set of a vector valued distribution.
Making use of a theorem by Dencker~\cite{Denck} 
and the equations of motion, we determine
the polarisation set of the twopoint function (Thm.~\ref{poltheorem}),
corresponding to a Hadamard state of the Dirac field. Along the
way, the propagation of singularities for the spinorial 
Klein-Gordon operator is obtained in Prop.~\ref{kgps}.

\medskip

On a Robertson-Walker spacetime and for free scalar fields, 
there exists the concept of `adiabatic states', which was
introduced a long time ago by Parker~\cite{parkerI} and 
put on a rigorous mathematical footing by 
L\"uders and Roberts~\cite{lr}. The main idea behind this concept is the following. 
If the scale factor $R(t)$ in the Robertson-Walker metric is constant in time, 
then there is an unambiguous notion of positive frequency solutions to the 
Klein-Gordon (KG) equation, and one can use these to define a ground state. 
If $R(t)$ is not constant, then no global ground state exists and 
the positive freqency solutions have to 
be determined dynamically off a given Cauchy surface (corresponding 
to the instant of time at which one wishes to define a vacuum-like state).  
They are usually found by a WKB-type ansatz and a subsequent iterative approximation 
process. Adiabatic states of order $n$ (at the time in question) are then the 
states defined from the positive frequency solutions obtained after $n$ approximation
steps. Recently, Junker~\cite{Junker} showed that the problem of finding the
`right' positive frequencies at an instant of time can 
be viewed as the problem to factorise the
KG-operator into positive and negative frequency 
parts near the Cauchy surface in question. 

In this work we present a construction (cf. Def.~\ref{adiabdef}) of adiabatic states
for Dirac fields on $(N+1)$-dimensional Robertson-Walker spacetimes, based
on a factorisation of the spinorial KG-operator near a Cauchy surface. 
While the construction is similar to the scalar case, there are 
also important differences; for example, special care needs 
to be taken in order to obtain a manifestly positive state. 
In Prop.~\ref{chargecon} we explain how the factorisation of the spinorial 
Klein-Gordon operator (and hence the split into 
positive and negative frequency solutions) is related
to the charge conjugation symmetry of the Dirac equation. 

In the scalar case,
L\"uders and Roberts have shown~\cite{lr} that the adiabatic states of
sufficiently high order are all locally quasi-equivalent and 
Junker~\cite{Junker} established that those of infinite order 
are of Hadamard type (which together implies that adiabatic states 
of sufficiently high order are locally quasiequivalent
to a Hadamard state\footnote{Originally, it had been claimed in 
\cite{Junker} that also adiabatic states of finite ordere were Hadamard. 
This has been corrected by the author of that paper in the meantime, 
cf.~\cite{JunkII}
}). In Sec.~IV of this work we show that these results also hold 
for our adiabatic states in the spin-1/2 case. Furthermore, we 
show that they correspond, in some sense, to a truncated Hadamard series 
and therefore allow for a point-splitting renormalisation of the stress tensor 
$T_{\mu\nu}$ in the same way as scalar fields; in other words, our
adiabatic states do not lead to infinite energy fluxes. 

Some of our results in the context of Robertson-Walker spacetimes can
be generalised to arbitrary globally hyperbolic spacetimes, for example
the construction of Hadamard states and a similar criterion for 
local quasiequivalence, based on the theory of pseudodifferential 
operators. For these and related issues we refer to a forthcoming 
paper. 

\medskip
\noindent
Concerning Thm.~\ref{poltheorem}: 
A proof of this was given 
in an earlier version of the present paper which was, however, 
unfortunately incorrect. The first correct proof was given 
by K.~Kratzert~\cite{kk}, see also~\cite{kkII}. These papers in 
turn built on earlier unpublished work by Radzikowski~\cite{rad}.
We are grateful to K.~Kratzert for communicating his results
to us and for pointing out the error in an earlier version. In the light
of his derivation of that result we were able to repair our earlier proof. 
It is included here since it is somewhat different from the proof 
in~\cite{kk, kkII}.

\section{The Dirac field on Robertson-Walker spacetimes}

The aim of this section is to recall the structure
of the Dirac equation on Robertson-Walker spacetimes. We assume that the
reader has some familiarity with the concept of spinors and the Dirac
equation in curved space, as described e.g. in ref.~\cite{Dim}. 

The homogeneous and isotropic spacetimes in $(N+1)$ dimensions are of
the form $M^\kappa = \mr \times \Sigma^\kappa$, the spatial section
$\Sigma^\kappa$ being the $N$-dimensional sphere $\ms^N$ for 
$\kappa = +1$, the Euclidean space $\mr^N$ for $\kappa = 0$ and 
the (real) hyperbolic 
space $\mh^N$ for $\kappa = -1$. The line-element on these 
spacetimes is 
\begin{eqnarray}\label{metric}
d s^2_\kappa = 
d t^2 - R^2(t)[d\theta^2 + f^2_\kappa(\theta) d\Omega_{N-1}^2] 
\end{eqnarray}
where 
\begin{eqnarray*}
f_\kappa(\theta) = 
\begin{cases}
\sin\theta & \text{for $\kappa = +1$}\\
\theta & \text{for $\kappa = 0$}\\
\sinh\theta & \text{for $\kappa = -1$}
\end{cases}
\end{eqnarray*}
and $d \Omega^2_{N-1}$ is the line-element on $\ms^{N-1}$. 
The above spacetimes are models for a 
closed, flat or hyperbolic universe with positive, zero or negative
curvature. 
We denote by $n^\mu\partial_\mu = \partial_t$ the future pointing
unit vector field normal to the Cauchy surfaces $\Sigma^\kappa$ and 
by $h^{\mu\nu} = g^{\mu\nu} - n^\mu n^\nu$ the induced 
(negative definite) metric on $\Sigma^\kappa$.
 
The spaces $\Sigma^\kappa$ are homogeneous for the groups 
$G^{+1} = Spin(N+1)$, 
$G^0 = Spin(N) \rtimes \mr^N$ and $G^{-1} = Spin(N,1)$ respectively, 
i.e. $\Sigma^\kappa = G^\kappa/K$, where $K = Spin(N)$. 
We shall omit the superscript $\kappa$ when  
not necessary and assume that $N$ is odd, $N \ge 3$ in order to 
simplify the exposition. 

\medskip

In order to bring out the 2 by 2 block matrix form of the Dirac equation
in RW-spacetimes, it is useful to define the associated vector bundles
(we view $G$ as a $K$ principal fibre bundle over $\Sigma$)
\begin{eqnarray*}
E^\tau = G \times_\tau \mc^{2^{(N-1)/2}}, \quad E^{\bar \tau}
= G \times_{\bar\tau} \mc^{2^{(N-1)/2}}, 
\end{eqnarray*}  
where $\tau$ is the fundamental representation of $K$ and
$\bar\tau$ the conjugate of that representation. They are 
related to each other by $Z\tau(k)Z^{-1} = \bar\tau(k)$ for all $k \in K$, 
where $Z$ is some unitary matrix. The spinor and
cospinor bundles (restricted to some Cauchy surface
$\Sigma(t) = \Sigma \times \{t\}$) are related to these bundles by
$DM \restriction \Sigma(t) = E^\tau \oplus E^\tau$ and $D^*M \restriction
\Sigma(t) = E^{\bar \tau} \oplus E^{\bar \tau}$.
Decomposing
\begin{eqnarray}
\label{decomp}
C^\infty(M, DM) \owns \psi = 
\left[
\begin{matrix}
\phi\\
\chi
\end{matrix}
\right], \quad \phi, \chi \in C^\infty(M, E^\tau), 
\end{eqnarray}
one can write the Dirac operator as the following 2 by 2 matrix
operator:
\begin{eqnarray*}
(i\dirac - m)\psi = \gamma^0
\left(i\partial_0 + \frac{iN}{2}\partial_0\log R 
-
\Bigg[
\begin{matrix}
m & i R^{-1}\widetilde{\dirac}\\
i R^{-1} \widetilde{\dirac} & -m
\end{matrix}
\Bigg]\right) 
\Bigg[
\begin{matrix}
\phi\\
\chi
\end{matrix}
\Bigg], 
\end{eqnarray*}
where
\begin{eqnarray*}
\gamma^0 = 
\left[
\begin{matrix}
1 & 0\\
0 & -1
\end{matrix}
\right].
\end{eqnarray*}
The operator $\widetilde\dirac$ is the Dirac operator on $E^\tau$,
defined {\it without} the scale factor $R$. The (generalised) 
eigenfunctions of this operator (normalised w.r.t. the 
natural inner product for sections in $E^\tau$),  
$\widetilde{\dirac} \chi_{\vec k s} = i sk\chi_{\vec k s}$,  
can be found in terms of special functions. 
The labels $(\vec k, s)$ mean
\begin{eqnarray*}
\vec k = (k, l, m) \quad \text{with}
\begin{cases}
k \in \mn + N/2, \,\, 0 \le l \le k-N/2, \,\, 
m = 0, \cdots, d_l &
\text{for $\kappa = +1,$}\\
k \in \mr_+,
\,\, l = 1, 2, \dots,
\,\, m = 0, \dots, d_l & 
\text{for $\kappa = 0, -1,$}
\end{cases}
\end{eqnarray*}
and $s = \pm 1$. $d_l$ is the degeneracy of the eigenvalue 
$l + (N-1)/2$ for the Dirac operator on $\ms^{N-1}$, 
given e.g. in~\cite{Trautman}. 
For an explicit representation of the functions 
$\chi_{\vec k s}$ we refer to \cite{CamporesiHiguchi} in the 
cases when $\kappa = -1, +1$. The case
$\kappa = 0$ is treated in the appendix. 

\medskip
\noindent
In order to diagonalise the Dirac Hamiltonian, one defines the spinors
\begin{eqnarray*}
u^+_{\vec k s} = R^{-N/2} \bU_{ks}
\left[
\begin{matrix}
\chi_{\vec k s}\\
0
\end{matrix}
\right], \qquad
u^-_{\vec k s} = R^{-N/2} \bU_{ks}
\left[
\begin{matrix}
0\\
\chi_{\vec k s}
\end{matrix}
\right]
\end{eqnarray*}
where
\begin{eqnarray}\label{UUU}
\bU_{ks} = \frac{1}{\sqrt{2}}
\left[
\begin{matrix}
\sqrt{1 + m/\omega_k} &  -s\sqrt{1 - m/\omega_k}\\
s\sqrt{1 - m/\omega_k} & \sqrt{1 + m/\omega_k}
\end{matrix}
\right]
\end{eqnarray}
is a unitary matrix and where $\omega_k = \sqrt{m^2 + k^2/R^2}$ are the instantaneous 
frequencies of the mode. The spinor fields 
$u^{\pm}_{\vec k s}$ form a complete set of generalised eigenfunctions
for the Hamiltonian 
\begin{eqnarray*}
H = -i\mslash n h^{\mu\nu}\gamma_\mu\nabla_\nu+ \mslash n m
= \Bigg[
\begin{matrix}
m & i R^{-1}\widetilde{\dirac}\\
i R^{-1} \widetilde{\dirac} & -m
\end{matrix}
\Bigg]
\end{eqnarray*}
and the helicity operator 
\begin{eqnarray}\label{heli}
\Xi = {\rm sign} (is^\mu\nabla_\mu) =
{\rm sign}
\left[
\begin{matrix}
i \widetilde\dirac & 0\\
0 & i \widetilde\dirac
\end{matrix}
\right], \quad
s^\mu = |g|^{1/2}
\epsilon^{\mu\nu\sigma\dots\rho}
n_\nu\gamma_\sigma\dots\gamma_\rho, 
\end{eqnarray}
with energy $\pm\omega_k$ and helicity $s$ at the corresponding instant of time,  
\begin{eqnarray}\label{eigenf}
H u^{\pm}_{\vec k s} = 
\pm \omega_k u^{\pm}_{\vec k s}, \quad
\Xi  u^{\pm}_{\vec k s} = 
s u^{\pm}_{\vec k s}. 
\end{eqnarray}
We next define the Dirac conjugate and the charge conjugate of a spinor or 
cospinor, which will be needed later on. Note that, 
as in any associated vector bundle, there is a one-to-one correspondence 
between sections $\chi$ in $E^\tau$ and smooth $\mc^{2^{(N-1)/2}}$-valued
functions $\chi^\wedge$ on $G$ such that 
$\chi^\wedge(gk) = \tau(k)^{-1}\chi^\wedge(g)$ (a similar identfication 
can be made for sections in $E^{\bar \tau}$). The charge conjugate
(denoted by $\psi^c$ or $C\psi$) resp. Dirac conjugate (
denoted by $\bar \psi$ or $\beta
\psi$) of a spinor $\psi$ with decomposition~\eqref{decomp} 
are then defined by 
\begin{eqnarray*}
\psi^c = 
\left[
\begin{matrix}
-(Z\phi^{\wedge\dagger})^\vee\\
(Z\chi^{\wedge\dagger})^\vee
\end{matrix}
\right], \quad
\bar\psi = 
\left[
\begin{matrix}
-(\chi^{\wedge\dagger})^\vee\\
(\phi^{\wedge\dagger})^\vee
\end{matrix}
\right]. 
\end{eqnarray*}
$\beta$ is an antilinear map from $DM$ to 
$D^*M$ and $C$ is an antilinear map from $DM$ to $DM$ for 
which $\{C, H\} = 0$. The charge resp. Dirac conjugate of 
a cospinor is defined in a similar way. The action of 
the symmetry group $G$ of the RW-spacetimes 
on spinors is defined as follows:
Firstly, one has an action $\widetilde{U}$ on sections in 
$E^\tau$, given by\footnote{
$\widetilde U$ is in fact the representation of $G$ induced
by the unitary representation $\tau$ of the closed compact subgroup $K$.
} 
\begin{eqnarray} \label{Gaction}
(\widetilde U(g)\chi)(\vec x) := (\chi^\wedge(g \, \cdot \, ))^\vee(\vec x). 
\end{eqnarray}
That group action then extends, by means of the 
isomorphisms $DM \restriction \Sigma(t) = E^\tau \oplus E^\tau$, 
to an action $U = \widetilde{U}\oplus \widetilde{U}$ on 
spinors over $M$. An action of $G$ on cospinors is defined
by $\overline{U}(g) = \beta U(g) \beta^{-1}$. 

An operator $B$ on $C^\infty_0(M, DM)$ is 
called {\rm isotropic} if it commutes with every $U(g), \, g \in G$. 
By abuse of notation, we will also use this term for operators defined 
only on a single Cauchy surface. Such operators have a mode decomposition
\begin{eqnarray}\label{ansatz}
\langle f_1, B f_2 \rangle_t = 
\int \rd\vec k \sum_s \sum_{pq}b_{ks}^{pq}
\overline{\tilde f_{1\vec k s}^p} 
\tilde f_{2\vec k s}^q, 
\end{eqnarray}
where $b_{ks}$ is some matrix valued function of the labels and 
$\tilde f^\pm_{\vec k s} = \langle f, u^\pm_{\vec k s} \rangle_t$,
the scalar product on a Cauchy surface $\Sigma(t)$ being defined by
\begin{eqnarray}\label{innprod}
\langle f_1, f_2 \rangle_t = 
\int_{\Sigma(t)}(\bar f_1 \gamma_\mu f_2)\,\rd S^\mu.
\end{eqnarray}
The correspondence $B \leftrightarrow b$ respects products and taking
hermitian adjoints, if these are well-defined. 
For a proof of the above facts we refer the reader
to~\cite{myphd}. For later use we mention that
$U$ can be viewed as a unitary representation of $G$ on 
$\cK_t=L^2(\Sigma(t), DM)$, the space of square integrable spinor 
fields w.r.t. the above inner product. 
A bounded isotropic operator $B$ on $\cK_t$ corresponds
to an essentially bounded function $b_{ks}$. 

The Dirac operator on a globally hyperbolic spacetime (such as
RW-spacetimes) has unique retared and advanced fundamental solutions
$\pgator_R$ and $\pgator_A$, see~\cite{Dim}, satisfying
\begin{eqnarray*}
(i\dirac - m)\pgator_A = \pgator_A(i\dirac - m) = \myid, \quad
(i\dirac - m)\pgator_R = \pgator_R(i\dirac - m) = \myid,   
\end{eqnarray*}
and (by $J^\pm$ we mean the causal future resp. past of a region in spacetime)
\begin{eqnarray*}
supp(\pgator_A f) \subset J^+(supp(f)), \quad
supp(\pgator_R f) \subset J^-(supp(f))
\end{eqnarray*}
for any compactly supported testspinor $f$. 
$\pgator = \pgator_A - \pgator_R$ is called the causal propagator.

\section{Local algebras for the Dirac field and invariant states}

\subsection{Local algebras of observables}
The Dirac field on globally hyperbolic manifolds can be quantised in a
straightforward manner. For convenience, we review the basic steps here,
details can be found in \cite{Dim}, which we follow closely. 
As above, let $\cK_t = L^2(\Sigma(t), DM)$ and $\cK'_t$
its topological dual, identified with $L^2(\Sigma(t), D^*M)$ (with 
the inner product defined in a similar way as in~\eqref{innprod}). 
The field algebra $F$ is the uniquely defined unital $C^*$-algebra 
$\car(\cK_t)$ generated by ``time $t$''--field operators 
$\Psi_t(f)$ and $\bar\Psi_t(h)$, smeared with square integrable 
spinor fields $f \in \cK_t$ resp. cospinor fields $h \in \cK'_t$, 
which satisfy the ``equal time $t$'' anti-commutation relations (CAR's)
\begin{eqnarray*}
\{ \Psi_t(f), \bar \Psi_t(h) \} = \langle \bar h, f \rangle_t
\myid, \quad
\bar \Psi_t(f)^* = \Psi_t(\bar f). 
\end{eqnarray*}
All other anti-commutators are trivial. One also defines $(N+1)$--smeared
field operators by $\Psi(f) = \Psi_t(\mslash S f \restriction \Sigma(t))$, 
where $f$ is now a compactly supported, smooth spinor field on $M$. 
In a similar way, one defines $\bar \Psi(h)$, where $h \in 
C^\infty_0(M, D^*M)$. The field $(N+1)$--smeared field operators 
satisfy by definition the field equations  
\begin{eqnarray*}
\bar \Psi((-i\dirac - m)h) = \Psi((i\dirac - m)f) = 0,  
\end{eqnarray*}
and the CAR's (see e.g. ref.~\cite{Dim}) 
\begin{eqnarray*}
\quad 
\{ \Psi(f), \bar\Psi(h) \} = i\pgator(h, f)\myid \quad \text{for all 
$f \in C_0^\infty(M, DM), h \in C^\infty_0(M, D^*M)$.}
\end{eqnarray*}
From this it follows at once that the definition of $F$ is 
in fact independent of the choice of Cauchy surface made above.  
The algebras of fields localised in a spacetime region 
$\cO$ are defined to be the $C^*$-algebras $F(\cO)$ 
generated by field operators smeared with
test functions supported in $\cO$. The algebras of observables localised in 
$\cO$ are given by $\cA(\cO) = F(\cO)^{even}$, where we mean the
subalgebras generated by products of an even number of fields. From the 
support properties of the causal propagator, one can easily deduce that
spacelike commutativity holds for the algebras of observables, 
\begin{eqnarray*}
[\cA(\cO), \cA(\cO')] = \{0\} \quad \text{if $\cO$ and $\cO'$ spacelike.}
\end{eqnarray*}
The group actions $U$ and $\overline U$ of $G$ on spinors resp. 
cospinors give rise, by standard results on the CAR, 
to an action by *-automorphisms 
$\alpha_g, \,\, g \in G$ on the field algebra $F$.
The action of these automorphisms on field operators is given by
\begin{eqnarray*}
\alpha_g \Psi_t(f) = \Psi_t(U(g)f), \quad \alpha_g \bar \Psi_t(h) = 
\bar\Psi_t(\overline{U}(g)h). 
\end{eqnarray*}
In the following, we will drop the subscript $t$ at the ``time $t$''--field
operators. This should cause no confusion, as it will be clear what is 
meant from the context. 

\subsection{Invariant, quasifree states}
A state $\omega$ on $\cA = F^{even}$ is said to be 
{\bf isotropic} if $\omega(X) = 
\omega(\alpha_g X)$ for all $g \in G$ and $X \in \cA$. It is said 
to be {\bf gauge invariant} and {\bf quasifree} if there exists an 
operator
$0 \le B \le \myid$ on $L^2(\Sigma(t), DM)$ such that 
\begin{eqnarray}\label{quasifreestate}
&\omega(\Psi(f_1) \dots \Psi(f_n)\bar\Psi(h_1) \dots \bar\Psi(h_m)) = 
\delta_{nm} {\rm det} \left(\langle
\bar h_i, B f_j \rangle \right)_{i,j = 1, \dots, n},\\ 
&h_i \in L^2(\Sigma(t), D^*M), \quad f_j \in L^2(\Sigma(t), DM)\nonumber.
\end{eqnarray}
The term ``gauge invariant'' refers to the fact that only monomials 
with the same number of $\Psi$ and $\bar \Psi$ fields have a nonzero 
expectation value in the state $\omega$. 
Clearly, a gauge invariant, quasifree state is isotropic 
if the corresponding operator $B$ is.

\medskip

One can easily show that the GNS-construction 
$(\pi_\omega, \cF_\omega, \Omega_\omega)$ of a gauge invariant 
quasifree state gives the following: 
$\cF_\omega$ is the antisymmetric Fock-space over $\cK_t \oplus \cK'_t$, 
$\Omega_\omega$ is the Fock-vacuum and the representation $\pi_\omega$
is  
\begin{eqnarray*}
\pi_\omega(\Psi(f)) = \hat a_+[B^{1/2}
(\pgator f) {\restriction\Sigma(t)}] + \hat a_-[\overline{(1-B)^{1/2}
(\pgator f)} {\restriction\Sigma(t)}]^*,
\quad f \in C^\infty_0(M, DM).  
\end{eqnarray*} 
Here, $\hat a_\pm$ are the destruction operators on $\cF_\omega$ for 
particles and antiparticles corresponding to the respective copies of 
$\cK_t$, which satisfy the usual anticommutation relations, 
\begin{eqnarray*}
\{\hat a_+(f_1)^*, \hat a_+(f_2)\} = \langle f_1, f_2 \rangle_t,  \quad 
\{\hat a_-(h_1)^*, \hat a_-(h_2)\} = \langle h_1, h_2 \rangle_t, 
\end{eqnarray*}
(all other anti-commutators vanish) 
and $\hat a_+(f)\Omega_\omega = \hat a_-(h)\Omega_\omega = 0$. 

\medskip

We next want to ask when two given quasifree, gauge invariant 
isotropic states are locally quasiequivalent.
Let $\omega_1$ and $\omega_2$ be two such states, 
corresponding to isotropic operators $B_1$ and $B_2$ (acting
on some Cauchy surface $\Sigma(t)$) with 
decompositions $b_1$ and $b_2$ as in Eq.~\eqref{ansatz}. 
\begin{thm}\label{qeq}
The states $\omega_1$ and $\omega_2$
are locally quasiequivalent provided
\begin{eqnarray*}
\esssup_{(k,s)} 
[(1 + |k|)^{N + \epsilon}\| b_{1ks} - b_{2ks} \|] < \infty
\end{eqnarray*}
(we mean the matrix norm in $\mc^2$) for some 
$\epsilon > 0$. 
\end{thm}
\begin{proof}
Let us choose a region $\cO$ of the form 
$D(\cC)$, where we mean the domain of dependence of some 
open subset $\cC$ with compact closure of the Cauchy surface $\Sigma(t)$, 
i.e. the set of all $x \in J^\pm(\cC)$ such that every
past resp. future directed timelike or null curve starting at $x$ hits $\cC$.
Let us first show that the restrictions of the states to a subalgebra
$\cA(\cO)$, $\cO = D(\cC)$ are quasiequivalent. 
Regions of this particular shape are convenient, because the algebras 
$\cA(\cO)$ are then isomorphic to the algebras $\car(\cK_\cC)$
constructed from the Hilbert space 
$\cK_\cC = L^2(\cC, DM)$ which is a closed
subspace of $\cK_t$. The restricted states 
$\omega_{1,2} \restriction \cO$
then correspond to the operators 
$E_\cC B_{1,2} E_\cC$ on $\cK_\cC$, where 
$E_\cC$ denotes the projection on this subspace. 
One then knows, by a well-known result of Powers and 
St\o rmer~\cite[Thm. 5.1]{ps}, that the states 
$\omega_1\restriction \cO$ and $\omega_2 \restriction \cO$ on the algebra 
$\car(\cK_\cC)$ are quasiequivalent if and only if  
\begin{eqnarray}\label{crit}
\|(E_\cC B_1 E_\cC)^{1/2} - (E_\cC B_2 E_\cC)^{1/2}\|_{\rm H.S.} 
&<& \infty, \nonumber\\
\|(E_\cC(I - B_1)E_\cC)^{1/2} - 
(E_\cC(I - B_2)E_\cC)^{1/2}\|_{\rm H.S.} &<& \infty, 
\end{eqnarray}
where we mean the Hilbert-Schmidt norm in $\cK_t$. Using the
Powers-St\o rmer inequality
$$\|A_1 - A_2\|_{\rm H.S.}^2 \le \|A_1^2 - A_2^2\|_{\rm tr}, $$
valid for any two bounded operators (we mean the trace norm), one 
concludes that Eqs.~\eqref{crit} hold provided
\begin{eqnarray}\label{traceclass}
\|E_\cC B_1 E_\cC - E_\cC B_2 E_\cC\|_{\rm tr} < \infty. 
\end{eqnarray}
As above, let $H$ be the Hamilton operator on $\cK_t$. 
Trivially, we can write ($p \in \mr$)
\begin{eqnarray*}
E_\cC B_1 E_\cC - E_\cC B_2 E_\cC = 
(E_\cC |H|^{-p/2})[|H|^{p/2}(B_1 - B_2)|H|^{p/2}](|H|^{-p/2} E_\cC).
\end{eqnarray*}
By assumption, the operator 
$|H|^{p/2}(B_1 - B_2)|H|^{p/2}$ is bounded for
$p \le N+\epsilon$, therefore Eq.~\eqref{traceclass} will hold if 
$E_\cC |H|^{-p/2}$ can be shown to be Hilbert-Schmidt for such a $p$, 
since the product of two Hilbert-Schmidt operators is in the 
trace class. To see this, let us pick an orthonormal basis 
$\{ f_n \}_{n \in \mn}$ of spinors in $\cK_\cC$. Then
\begin{eqnarray}
\label{*}
\|E_\cC |H|^{-p/2}\|_{\rm H.S.}^2 &=& 
\sum_n \langle f_n, E_\cC |H|^{-p} E_\cC f_n \rangle_t\nonumber\\ 
&=&\sum_n \int \rd \vec k \sum_{qs} (k^2/R^2 + m^2)^{-p/2}
|\tilde f_{n\,\vec k s}^q|^2.
\end{eqnarray}
In order to estimate the r.h.s. of Eq.~\eqref{*}, we
exchange the $\rd \vec k$--integration and the 
summation over $n$ (this is justified, because the resulting
expression turns out to be absolutely convergent). We have, 
\begin{eqnarray*}
\sum_{n} |\tilde f^q_{n\,\vec k s} |^2
= \int_\cC u^{q}_{\vec k s}(\vec x)^\dagger
u^{q}_{\vec k s}(\vec x) R^N \rd^N\vec x.
\end{eqnarray*}
The sum over $q, s, \vec k$ at fixed $k$ of the integrand 
is independent of $\vec x$ and equal to twice the spectral 
function $P_N(k)$, defined by
\begin{eqnarray*}
P_N(k) = \sum_{lm}
\chi_{klms}(0)^\dagger \chi_{klms}(0). 
\end{eqnarray*}
therefore we have found
\begin{eqnarray*}
\|E_\cC |H|^{-p/2}\|_{\rm H.S.}^2  = 2\,{\rm vol}(\cC) \int 
(k^2/R^2 + m^2)^{-p/2} P_N(k)\rd k.
\end{eqnarray*}
The spectral function $P_N$ is given by 
the following expressions
\begin{eqnarray*}
P_N(k) = 
\begin{cases}
2^{(N-1)/2}\frac{(N+k-1)!}{k!(N-1)!} & \text{for $\kappa = +1$,}\\
\frac{k^{N-1}}{2^{N/2}\,{\rm vol}(\ms^{N-1})\Gamma(N/2)^2} & \text{for 
$\kappa = 0$,}\\
\frac{\pi}{2^{2N-4}}\Big\vert\frac{\Gamma(N/2+ik)}{
\Gamma(N/2)\Gamma(1/2 + ik)} \Big\vert^2 & 
\text{for $\kappa = -1$.}  
\end{cases}
\end{eqnarray*}
For $\kappa = -1, +1$ a derivation of these may be found
in~\cite{CamporesiHiguchi}, the expression for $\kappa = 0$ is derived
in the Appendix. $P_N(k)$ grows as $k^{N-1}$ for large $k$ for
for all the homogeneous spaces $\Sigma = \mr^N, \ms^N, \mh^N$, ensuring 
that $E_\cC |H|^{-p/2}$ is Hilbert-Schmidt for 
$p = N+\epsilon$. We have therefore shown that $\omega_1 \restriction 
\cO$ is quasiequivalent to $\omega_2 \restriction \cO$ for any 
set $\cO$ of the form $D(\cC)$. 
Since it is enough to verify local quasiequivalence 
on a cofinal set of open subsets (such as the set of regions of the 
type $D(\cC)$), this then proves the theorem. 
\end{proof}

\section{General properties of Hadamard states for the
Dirac field}

The plan of this section is as follows: We first give a microlocal 
definition of Hadamard states for Dirac quantum fields. After that
we determine the polarisation set of the twopoint function of 
such states.  Finally, we 
explain how our microlocal notion of Hadamard states 
is related to pre-existing 
notions of Hadamard states for the Dirac fields based on the
short-distance behaviour of the twopoint functions. The 
definitions and results in this section apply to the case 
of a general, globally hyperbolic spacetime. 
The reader not familiar with the technical ingredients of the 
definition will find some notation and results from microlocal
analysis in the appendix. 

The spatio-temporal twopoint functions of a state $\omega$ are denoted
by
\begin{eqnarray*}
\mslash G^{(+)}(h, f) := \omega(\Psi(f) \bar\Psi(h)),
\quad  
\mslash G^{(-)}(h, f) := \omega(\bar\Psi(h)\Psi(f)), 
\end{eqnarray*}
where $f \in C^\infty_0(M,DM)$ and $h \in C^\infty_0(M,D^*M)$. 
They are assumed to be distributions. We introduce the 
following (standard) notation: elements in $T^*_x M$ are
denoted by $(x, \xi)$. We write
$(x_1, \xi_1) \sim (x_2, \xi_2)$, if
$x_1$ and $x_2$ can be joined by a null-geodesic $c$ such that 
$\xi_1 = \dot c(0)$ and $\xi_2 = \dot c(1)$, or
if $(x_1, \xi_1) = (x_2, \xi_2)$ and 
$\xi_1$ null. We write $x_1 \succ x_2$ resp. 
$x_1 \prec x_2$ if the point $x_1$ comes after or before $x_2$ according to 
the parameter on this curve. 
Moreover, we shall write $\xi
\triangleright 0$ if $\xi$ is future-directed and $\xi \triangleleft 0$ if
it is past-directed.

\begin{defn}\label{Hadamardcond} 
A quasifree state $\omega$ for the Dirac field is said to be `Hadamard' if 
\begin{multline} 
\label{77}
\WF'(\mslash G^{(\pm)}) = \{
(x_1, \xi_1, x_2, \xi_2) \in T^*M \backslash \{0\} \times 
T^*M \backslash \{0\} \mid \\
(x_1, \xi_1) \sim  (x_2, \xi_2), \quad \xi_1 
\triangleright (\triangleleft) \,
0 \}
=: C^{(\pm)}
\end{multline}
\end{defn}
\begin{remark}
It follows from what we say in the proof of Thm.~\ref{adahad}
that the apparently weaker condition $\WF'(G^{(\pm)}) \subset
C^{(\pm)}$ already implies equality of these sets. 
\end{remark}

The conditions on the wave front set of $\mslash G^{(+)}$ and
$\mslash G^{(-)}$ are in general independent, i.e., there 
exist states for which 
$\mslash G^{(+)}$ has the desired wave front set but for 
which $\mslash G^{(-)}$ has not. However, it is easy to see
that this cannot happen for states $\omega$ which are 
invariant under charge conjugation.
By this we mean that 
$\omega \circ \alpha_c = \omega$, where $\alpha_c$ is 
the $^*$-automorphism defined by 
\begin{eqnarray*}
\alpha_c\Psi(f) = \bar \Psi(\beta f^c), \quad
\alpha_c \bar \Psi(h)  = - \Psi(\beta h^c).
\end{eqnarray*}
Below (Thm.~\ref{hadachar}) we prove that the above condition on the wave
front sets of {\it both} twopoint functions implies that their 
singular behaviour in position space is that of a Hadamard 
fundamental solution. This result will in general 
{\it not hold} if only {\it one} twopoint function 
has the required wave front set.  

\medskip
We now analyse in more detail the microlocal singularity structure of the 
twopoint functions 
by applying to them the propagation of singularities 
theorem Thm.~\ref{propsing}, 
taking $P$ in that theorem 
to be the spinor Klein-Gordon (KG) operator, given by 
\begin{eqnarray*}
P = \square + \frac{1}{4} R + m^2,  
\end{eqnarray*} 
where $R$ is the curvature scalar. This theorem is 
applicable because 
\begin{eqnarray}
\label{eqomo} 
(P \otimes \myid)\slom^{(\pm)} = 
(\myid \otimes P^t)\slom^{(\pm)} = 0, 
\end{eqnarray}
which is in turn a direct consequence of the Lichnerowicz 
identity
\begin{eqnarray*}
P = (i\dirac - m)(-i\dirac - m)
\end{eqnarray*}
and the fact that $\slom^{(\pm)}$ satisfy the 
Dirac equation. By  
$^t$ we mean the transpose of an operator acting in 
the dual bundle (the bundle $D^*M$ in the case at hand), obtained 
from the natural pairing between spinor and cospinor fields on $M$. 
(The assumption in Thm.~\ref{propsing} that $P$ be of principal
type (cf.~Def.~\ref{realprincipal}) is fulfilled, 
because $P$ has a metric principal symbol 
$p_0(x, \xi) = -g^{\mu\nu}(x)\xi_\mu\xi_\nu$.) In order to apply
Thm.~\ref{propsing}, we must calculate the Dencker connection $D_P$
associated with $P$, defined in Eq.~\eqref{dencon}. In the case
at hand, $D_P$ naturally acts in the vector bundle 
$\pi^*(DM \otimes \Omega^{1/2}) \restriction {\cal Q}_P$, where
$\Omega^{1/2}$ is the line bundle over $M$ of half-densities, 
$\pi: T^*M \to M$ is the projection map and 
\begin{eqnarray*}
\cQ_P = \{(x, \xi) \mid g^{\mu\nu}(x)\xi_\mu\xi_\nu = 0\} \subset T^*M.     
\end{eqnarray*}
\begin{prop}\label{kgps}
The Dencker connection $D_P$ for the operator $P$ is the partial
connection in the pull-back of the vector bundle $DM \times \Omega^{1/2}$
to $\cQ_P \subset T^*M$ given by
\begin{eqnarray*}
D_P = {\rm i}_{X_{p_0}} \circ \pi^*( \nabla + \frac{1}{4} d \log |g|). 
\end{eqnarray*}
Here, $X_{p_0}$ is the Hamiltonian
vector field over $T^*M$ corresponding to $p_0(x, \xi)$, 
$\frac{1}{4}d\log |g|$ is the natural connection in $\Omega^{1/2}$ and
${\rm i}_{X_{p_0}}$ is the insertion operator.
\end{prop}
\begin{proof}
The Dencker connection can be calculated from Eq.~\eqref{dencon}, 
taking $\tilde p_0 = \myid$ (this means that $q = p_0$ in that 
formula). Introducing an orthonormal frame 
$e^a_\mu \rd x^\mu$ for the metric tensor, 
$\eta^{ab} = g^{\mu\nu}e^a_\mu e^b_\nu$, and 
and going to local coordinates, we find 
\begin{eqnarray*}
\{\tilde p_0, p_0 \} = 0, \quad 
p^s(x, \xi) = 2iS_\mu(x)\xi^\mu - ie^a_\mu(x)\partial_\nu e^\mu_a(x)\xi^\nu
\end{eqnarray*}
and 
\begin{eqnarray*}
X_{p_0}(x, \xi) = -2\xi^\mu\frac{\partial}{\partial x^\mu} + 
2 e^a_\sigma(x)\partial_\mu e_{a\nu}(x)
\xi^\nu\xi^\sigma
\frac{\partial}{\partial \xi_\mu}, 
\end{eqnarray*}
where $S_\mu = \frac{1}{8} [\gamma^a, \gamma^b]
e^\nu_a \nabla_\mu e_{\nu b}$ is the spinor connection. 
According to Eq.~\eqref{dencon}, this means that 
\begin{eqnarray*}
D_P = -2\xi^\mu\frac{\partial}{\partial x^\mu} + 
2 e^a_\sigma(x)\partial_\mu e_{a\nu}(x)
- 2S_\mu(x)\xi^\mu + e^a_\mu(x)\partial_\nu e^\mu_a(x)\xi^\nu.
\end{eqnarray*}
That this expression agrees with the 
geometric expression for $D_P$ in the statement of this proposition
now immediately follows, because
$\nabla_\mu = \partial_\mu + S_\mu$,  
$\frac{1}{4}\partial_\mu \log|g| = -\frac{1}{2} e^a_\nu \partial_\mu
e_a^\nu$, and because
the differential of the projection map is given by 
$d\pi(\tfrac{\partial}{\partial x^\mu}) = 
\tfrac{\partial}{\partial x^\mu}, \,
d\pi(\tfrac{\partial}{\partial \xi^\mu}) = 0$. 
\end{proof}

Having derived this tool, we can establish the following result on
the polarisation set of the twopoint functions of a Hadamard state.

\begin{thm}\label{poltheorem}
Any Hadamard state has the following polarisation set:
\begin{multline*}
\Pol'(\slom^{(\pm)}) = \{ (x_1, \xi_1, x_2, \xi_2, w) \mid 
(x_1, \xi_1, x_2, \xi_2) \in C^{(\pm)}, w \in 
D_{x_1} M \otimes D_{x_2} M,\\
\text{and ${w^A}_{C'}{\cI(x_1, x_2)_B}^{C'} = \lambda {\mslash {\xi_1}
^A}_{B}$ for some  $\lambda \in \mc$}
\}.
\end{multline*}
Here, unprimed spinor indices refer to the point $x_1$, whereas 
primed ones correspond to $x_2$ and $\cI(x_1, x_2)$
is the bispinor of parallel transport in the bundle $DM$ along a 
null geodesic joining $x_1$ and $x_2$. The sets $C^{(\pm)}$ 
had been defined in Eq.~\eqref{Hadamardcond}.  
\end{thm}
\begin{proof}
We aim at using the propagation of singularities theorem, 
Thm.~\ref{propsing}, combined with a deformation argument due to 
Fulling, Narcowich and Wald \cite{fnw}, first applied in a similar context
in \cite{koeh}. 

By Thm.~\ref{propsing} and Eq.~\eqref{eqomo}, 
the polarisation set of $\slom^{(\pm)}$ must be a union of Hamilton orbits
corresponding to the operators $P \otimes \myid$ and $\myid \otimes P^t$. 
By the Prop.~\ref{kgps}, sections over $\cQ_P$, annihilated by $D_P$ are 
pull-backs to $T^*M$ of sections in $DM$ over null-geodesics which 
are parallel with respect to $\nabla$.  Therefore, two elements 
$(x_1, \xi_1, x_2, \xi_2, w)$ and
$(x_1', \xi_1', x_2', \xi_2', w')$ of 
$\pi^* (DM \boxtimes D^*M)$ 
are in the same Hamiltonian orbit if 
\begin{eqnarray*}
(x_1, \xi_1) \sim (x_2, \xi_2), \quad (x_1', \xi_1') \sim (x_2',
\xi_2'),\\
\lambda \cdot {w^A}_B = {\cI(x_1, x_1')^A}_{B'}
{w^{\prime B'}}_{C'}{\cI(x_2, x_2')^{C'}}_B, 
\end{eqnarray*}
for some $\lambda \in \mc$. 
Now let $x \in M$ and $\Sigma$ be a Cauchy 
surface of $M$ through that point. Then there is there is a 
convex normal $U$ of $x$ and a convex normal
neighbourhood $V$ of $\Sigma$, containing $U$,  
such that there is another spacetime $(\hat M, \hat 
g)$ with Cauchy surface $\hat \Sigma$ and a corresponding causal
normal neighbourhood $\hat V$ with the properties that: (a) $(V, g)$
is isometric to $(\hat V, \hat g)$ and (b) $\hat M$ contains a
Cauchy surface $\hat \Sigma_1$ and a  
a convex flat neighbourhood $\hat U_1$ contained in a convex normal 
neighbourhood $\hat V_1$ of $\hat\Sigma_1$ such 
that $D(\hat U_1) \supset \hat U$ (we mean the domain of dependence), 
where $\hat U$ corresponds to $U$ under the isometry. By the propagation of 
singularities theorem, it will be enough to show that $ 
\slom^{(\pm)}\restriction U \times U$ has the desired polarisation set, 
because any pair of null related points can be transported 
along a null geodesic into a region of that kind. Let 
$\hat \slom^{(\pm)}\restriction \hat V \times \hat V$ be 
the pull-back of the twopoint functions to the deformed spacetime $(\hat
V, \hat g)$. By the propagation of singularities theorem and the
equations of motion on the deformed spacetime it will induce a Hadamard 
distribution on all of $\hat M$. Furthermore 
$\hat \slom^{(\pm)}\restriction \hat
U \times \hat U $ will have the required polarisation set if 
$\hat\slom^{(\pm)}\restriction{\hat U_1 \times \hat U_1}$ has, 
again by the propagation of singularities theorem. But $\hat U_1
\subset \hat M$ is contained in a flat portion of spacetime, so
effectively our theorem has to be shown for Minkowski space only. 
 
\medskip
 
So let $\slom^{(\pm)}_{\rm mink}$ be twopoint functions of a 
Hadamard state in Minkowski space. By our Thm.~\ref{hadachar}, 
all Hadamard states differ by a smooth piece only, so we might restrict
attention to the vacuum in Minkowski space, 
\begin{eqnarray*}
\slom^{(\pm)}_{\rm mink} = \pm (i\mslash\partial + m) 
\Lambda^{(\pm)}_{\rm mink}, 
\end{eqnarray*}
where $\Lambda^{(\pm)}_{\rm mink}$ are the ordinary positive resp. negative 
frequency twopoint functions for the KG-operator $\eta^{\mu\nu}\partial_\mu 
\partial_\nu + m^2$ in flat space. 
It is not difficult to see from the definition of the polarisation
set and the definition of $\slom^{(\pm)}_{\rm mink}$ that one must have
\begin{multline}
\label{prelim}
\Pol(\slom^{(\pm)}_{\rm mink}) \subset
\{(x_1, \xi, x_2, -\xi, w) \mid  (x_1, \xi, x_2, -\xi) \in 
\WF(\Lambda^{(\pm)}_{\rm mink}),\\
\text{$w = c\myid + \mslash\beta$ 
for some $c \in \mc$ and $\beta \in \mc^{N+1}$}  
\}. 
\end{multline}
Now
\begin{eqnarray*}
[(i\mslash\partial - m) \otimes \myid] \slom_{\rm mink}^{(\pm)} = 
\slom_{\rm mink}^{(\pm)}[\myid \otimes (-i \mslash\partial - m)] = 0, 
\end{eqnarray*}
therefore, using that $\mslash \xi$ is a principal symbol of 
$i\mslash\partial + m$, one can conclude (by the definition of the polarisation set) 
that 
\begin{eqnarray*}
\mslash \xi w = w \mslash \xi = 0,
\end{eqnarray*} 
where $(x_1, \xi, x_2, -\xi) \in \WF(\slom^{(\pm)}_{\rm mink})$.
Since the form of $w$ is already restricted by Eq.~\eqref{prelim}, 
it is easy to see that these equations imply
$\eta^{\mu\nu}\xi_\mu\beta_\nu = 0$ and $c = 0$. From the 
equation $\mslash\beta\mslash\xi = 0$ it follows
\begin{eqnarray*}
\tr(
\gamma^{\alpha_1}\dots
\gamma^{\alpha_{N-1}}\gamma^{N+2}
\epsilon_{\alpha_1\dots\alpha_{N-1}\sigma\rho}\mslash\beta\mslash\xi )
= 0
\end{eqnarray*}
where $\epsilon_{\mu \dots \nu}$ is the totally antisymmetric tensor
in $N + 1$ dimensions and
\begin{eqnarray*}
\gamma^{N+2} = \gamma^0 \dots \gamma^d.
\end{eqnarray*}
Using standard identities for traces of gamma matrices and
\begin{eqnarray*}
\epsilon^{\mu\nu\alpha_1 \dots \alpha_{N-1}}
\epsilon_{\alpha_1 \dots \alpha_{N-1}\sigma\rho}
= \delta^\mu_\sigma \delta^\nu_\rho - 
\delta^\mu_\rho\delta^\nu_\sigma
\end{eqnarray*}
we find that $\xi_\rho\beta_\sigma - \xi_\sigma\beta_\rho = 0$. Since $\xi \neq 0$, this implies
$\mslash\beta = w = \lambda \cdot \mslash \xi$, which in Minkowski space 
is just the condition on the polarisation that was claimed. 
\end{proof}

In \cite{ver, koeh}, the authors give a definition of 
the Hadamard condition for Dirac fields in terms of the singular
behaviour of the associated twopoint functions in position space. 
We now investigate the relation between the two definitions.
A local version of the definition of \cite{ver, koeh} may be 
stated as follows. Let $\cO$ be a convex normal neighbourhood in $M$.
On $\cO$, one defines the bidistributions 
\begin{eqnarray*}
H_n^{(\pm)}(x_1, x_2) = 
\sum_{j = 1}^{(N-1)/2} U_j(x_1,
x_2)\sigma_{\pm\epsilon}^{-j} + \sum_{j = 0}^{n + 1} 
V_j(x_1, x_2) \sigma^j \log \sigma_{\pm\epsilon} 
\end{eqnarray*}
where $\sigma$ is the signed squared geodesic distance between 
the points $x_1, x_2 \in \cO$, 
\begin{eqnarray*}
\sigma_\epsilon(x_1, x_2) = \sigma(x_1, x_2) + 2i\epsilon (t(x_2) - 
t(x_1)) + \epsilon^2, \quad \epsilon > 0 
\end{eqnarray*}
and $t$ is some global time function. As usual, we mean the bidistribution 
obtained by smearing with smooth spinor fields 
first and then taking $\epsilon$ to zero. The bispinors $U_j, V_j$ are
determined recursively by the $N+1$ dimensional analogue of the Hadamard
transport equations \cite{moretti} for the spinorial KG-operator
$P$ and depend only the geometry of the spacetime in $\cO$.
By construction, 
\begin{eqnarray*}
(P \otimes \myid)H_n^{(\pm)}  = (\myid \otimes P^t) H_n^{(\pm)} = 0
\quad \text{mod $C^n$}
\end{eqnarray*}
on $\cO$. The local version of the global Hadamard 
definition\footnote{
Not all points in $M$ may be connected by a unique geodesic line, 
therefore the definition needs to be refined if one also 
wants to exclude spacelike singularities. We refer to 
\cite{kaywald} for details.}
given in \cite{ver, koeh} is as follows: A 
state is said to be locally Hadamard if its 
associated twopoint function satisfies 
\begin{eqnarray}\label{mparametrix}
\mslash G^{(\pm)} = \pm (-i\dirac-m) H^{(\pm)}_n 
\quad \text{modulo $C^n$ on $\cO$ for all $n$.}    
\end{eqnarray} 
In fact, it is only necessary to require this equation
for `$+$' (as the authors do in ref.~\cite{ver, koeh}), 
since it automatically follows from this that 
\begin{eqnarray*}
\slom^{(-)} &=& i(-i\dirac - m)E - \slom^{(+)}\\
&=& (-i\dirac - m)(iE - H^{(+)}_n) \quad \text{mod $C^n$}\\
&=& -(-i\dirac - m)H^{(-)}_n \quad \text{mod $C^n$,}
\end{eqnarray*}
where we have used that 
$$
H^{(+)}_n - H^{(-)}_n = iE \quad \text{mod $C^{n+1}$,}
$$
and where $E$ is the causal propagator for $P$. 
This follows from the fact~\cite{friedl} 
that $E$ contains the same coefficients
$U_j, V_j$ as $H^{(\pm)}_n$, multiplied only by different singular pieces, which, 
as it is well-known in QFT, may be combined to give 
the above identity between the different 
types of propagators.

We now show that Eq.~\eqref{mparametrix} 
follows from our microlocal definition. 
This observation has been made first by
Radzikowski~\cite{Radzikowski}
for the 
scalar field and his proof can be adjusted to the spinor case as well, 
although the situation is more complicated. 

\begin{thm}\label{hadachar}
Let $\omega$ be a Hadamard state in the microlocal sense of 
Def.~\ref{Hadamardcond}. Then its twopoint functions
also satisfy Eq.~\eqref{mparametrix}. 
\end{thm}
\begin{remark}
It can also be shown that the global version of the Hadamard 
condition~\eqref{mparametrix} (as for example spelled out 
in~\cite{ver, koeh}) implies the microlocal Hadamard condition. 
A proof of this is almost identical to the proof of the 
corresponding statement for the scalar case. We refer to
refs.~\cite{koeh,kk} for details.
\end{remark}
\begin{proof}
Let us denote by $E_A, E_R, E_F, E_{\bar F}$ the advanced, retarded, 
Feynman, anti-Feynman parametrices of the spinorial Klein-Gordon
operator $P$.  They are known to be determined (modulo  
$C^\infty$) by the equations 
\begin{eqnarray*}
P E_j = \myid \quad \text{mod $C^\infty, \, 
j = A, R, F, \bar F$}
\end{eqnarray*}
and their wave front sets~\cite[Thm. 6.5.3]{hodu}. For the Feynman and
anti-Feynman parametrices, these wave front sets read:
\begin{multline}
\label{wfs}
\WF'(E_F) = \{
(x_1, \xi_1, x_2, \xi_2)\mid  
(x_1, \xi_1) \sim  (x_2, \xi_2),\\
\text{  
$\xi_1 \triangleright 0$ if $x_1 \succ x_2$,   
$\xi_1 \triangleleft 0$ if $x_1 \prec x_2$ }
\}
\end{multline}
and 
\begin{multline}
\WF'(E_{\bar F}) = \{
(x_1, \xi_1, x_2, \xi_2) \mid  
(x_1, \xi_1) \sim  (x_2, \xi_2),\\
\text{
$\xi_1 \triangleright 0$ if $x_1 \prec x_2$, 
$\xi_1 \triangleleft 0$ if $x_1 \succ x_2$ }
\}.
\end{multline}
Actually, in \cite[Thm. 6.5.3]{hodu} only the 
case of a scalar operator with metric principal part is treated. 
Inspection of the proof however shows 
that it may be extended to operators with metric 
principal part acting in vector bundles such as $P$. 
We also need the advanced, retarded, Feynman and anti-Feynman parametrices for the 
Dirac operator, given by 
\begin{eqnarray*}
\mslash S_j = (-i\dirac - m) E_j, \quad j = A, R, F, \bar F.
\end{eqnarray*}
By the anticommutation relations, one infers that 
\begin{eqnarray*}
\slom^{(+)} + \slom^{(-)} = i\mslash S =
i(\mslash S_A - \mslash S_R).
\end{eqnarray*}
Let us define 
\begin{eqnarray}\label{feynmanpro}
\slom_F &:=& i\mslash
\slom^{(+)} + \mslash S_A = -i\slom^{(-)} + \mslash S_R\nonumber\\
\slom_{\bar F} &:=& - i\mslash
\slom^{(+)} + \mslash S_R = i\slom^{(-)} + \mslash S_A.
\end{eqnarray}
Our aim is now to prove that 
\begin{eqnarray*}
\slom_F = \mslash S_F, \quad 
\slom_{\bar F} = \mslash S_{\bar F}
\end{eqnarray*}
modulo a smooth kernel. 
To this end, we first show that 
\begin{eqnarray}
\label{includ}
\WF'(\slom_F) \subset \WF'(E_F), \quad 
\WF'(\slom_{\bar F}) \subset \WF'(E_{\bar F}).
\end{eqnarray}
In order to see why this must be true, consider a point
$(x_1, \xi_1, x_2, \xi_2)$ in wave front set of 
$\slom_F$ such that $x_1 \notin J^-(x_2)$. 
Then, because $\mslash S_A$ must be zero for 
such points by the support properties of $E_A$,
it must hold that $(x_1, \xi_1, x_2, \xi_2) \in 
\WF'(\slom^{(+)})$. Since (by the microlocal Hadamard 
condition, Eq.~\eqref{77}) $\WF'(\slom^{(+)}) \subset C^{(+)}$ we find that 
$(x_1, \xi_1, x_2, \xi_2)$ can be 
in the wave front set of $\slom_F$ if and only if 
$\xi_1 \triangleright 0$. A
similar reasoning can be applied for $x_1 \in J^-(x_2)$, this time using
the representation $\slom_F = -i\slom^{(-)} + \mslash S_R$ and
exploiting the microlocal Hadamard condition, 
Eq.~\eqref{77}, for $\slom^{(-)}$.
Altogether one concludes from this that 
$(x_1, \xi_1, x_2, \xi_2)$ is in the wave front 
set of $\slom_F$ if and only if $(x_1, \xi_1) \sim (x_2, \xi_2)$
and $\xi_1 \triangleright 0$ for $x_1 \succ x_2$ 
resp. $\xi_1 \triangleleft 0$ if $x_1
\prec x_2$, which is just the set Eq.~\eqref{wfs}.
We have therefore shown the first inclusion in Eq.~\eqref{includ}.
The second inclusion is treated in just the same way. 

Now, by definition, we have $\slom_F + \slom_{\bar F} =
\mslash S_A + \mslash S_R$. Applying
the operator $-i\dirac - m$ to the 
relation~\cite[II, Eq. 6.6.1]{hodu}
\begin{eqnarray*}
E_F + E_{\bar F} = E_A + E_R
\quad\text{mod $C^\infty$,}
\end{eqnarray*}
we find from this that $\slom_F - \mslash S_F = 
\slom_{\bar F} - \mslash S_{\bar F}$ modulo smooth and hence that 
\begin{eqnarray}
\label{ausgleich}
\WF'(\slom_F - \mslash S_F) = \WF'(\slom_{\bar F} - \mslash
S_{\bar F}). 
\end{eqnarray}
We had already shown that 
$\WF'(\slom_F) \subset \WF'(E_F)$ and we also have 
$\WF'(\mslash S_F) = \WF'((-i\dirac-m)E_F) \subset \WF'(E_F)$ 
(since the wave front set of a distribution 
cannot become larger when acting upon it 
with a differential operator), so the set on the 
left had side of this equation is contained in the set
$\WF'(E_F)$. By the same arguments
the set on the right hand side Eq.~\eqref{ausgleich}
must be contained in $\WF'(E_{\bar F})$. It therefore 
trivially follows that 
\begin{eqnarray*}
\WF'(\slom_F - \mslash S_F) \subset \Delta_{\cQ_P}, \quad
\WF'(\slom_{\bar F} - \mslash S_{\bar F}) \subset \Delta_{\cQ_P}, 
\end{eqnarray*} 
where 
\begin{eqnarray*}
\Delta_{\cQ_P} = \WF'(E_F) \cap \WF'(E_{\bar F}) = 
\{(x, \xi, x, \xi) \mid g^{\mu\nu}(x)\xi_\mu\xi_\nu = 0\}.
\end{eqnarray*}
Now by definition
\begin{eqnarray*}
(P \otimes \myid)(\slom_F - \mslash S_F) &=& 
(\myid \otimes P^t)(\slom_F - \mslash S_F) = 0
\end{eqnarray*}
modulo smooth. We are thus in a position to use the propagation of singularities
theorem and we conclude that polarisation sets of the distribution
$\slom_F - \mslash S_F$ (and hence the wave
front set) must be  a union of Hamiltonian orbits for the operator $P$. Using the same
arguments as in the proof of the preceding theorem, it is then easy to 
show that if $(x, \xi, x, \xi)$ is in $\WF'(\slom_F - \mslash S_F)$, 
then this set must also contain nonzero vectors away from the diagonal, 
a contradiction. Hence $\WF'(\slom_F - \mslash S_F) = \emptyset$, i.e.,  
\begin{eqnarray*}
\slom_F - \mslash S_F \in C^\infty, 
\end{eqnarray*} 
as we wanted to show. Inserting this into
Eq.~\eqref{feynmanpro} one gets
\begin{eqnarray}\label{++}
i \slom^{(+)} &=& \slom_F - \mslash S_A = 
(-i\dirac - m)(E_F - E_A) \quad \text{mod $C^\infty$.}
\end{eqnarray}  
It can be extracted from the analysis of the propagators in 
\cite{friedl,gu} that 
\begin{eqnarray*}
E_F -  E_A &=& iH_n^{(+)} \quad 
\text{mod $C^{n+1}.$}
\end{eqnarray*}
This holds because all propagators have the same structure
as $H_n^{(\pm)}$, i.e. the same functions $U_j, V_j$ 
multiply different singular parts. These can be combined 
to give the above equation. Combining this equation with 
Eq.~\eqref{++} then proves the theorem. 
\end{proof}

\section{Adiabatic states}\label{adiabaticstates}

\subsection{Definition of adiabatic states}

For the remainder of this work, we restrict attention to Dirac fields over
RW-spacetimes. It will be clearly indicated if a result has a wider range of validity. 

We begin by constructing adiabatic states for 
the Dirac fields on RW-spacetimes. By term ``adiabatic'' we mean that 
any of these states should give a 
reasonable mathematical description of the concept of `empty space' in the 
very small, i.e. in spacetime regions which are very small compared to the
curvature radius. 
The main ingredient in the construction is a factorisation of the spinorial 
Klein-Gordon operator into positive and negative frequency 
parts, which has been considered before in~\cite{Junker}, 
for the spin-0 case. (Such a factorisation is possible on every globally hyperbolic 
spacetime with a compact Cauchy surface.)   
Let us outline our construction. 
Suppose one has constructed a `Gau\ss ian foliation' 
$I \owns t \mapsto \Sigma(t)$
of a neighbourhood of a globally hyperbolic spacetime
$M$ by smooth Cauchy surfaces $\Sigma(t)$, 
$I = [t_0, t_1]$ denoting some time interval.
By `Gau\ss ian', we mean that the vector field $\partial_t = n^\mu\partial_\mu$ 
associated to this foliation is geodesic and orthonormal to 
the Cauchy surfaces, in other words 
$\nabla_\nu n_\mu - \nabla_\mu n_\nu = 0$ and 
$n^\mu\nabla_\mu n^\nu = 0$. It is then possible to make the 
decomposition ($P$ is the spinorial KG-operator)
\begin{eqnarray}\label{factorisation}
P =
-(in^\mu\nabla_\mu + iK + T)(in^\mu\nabla_\mu - T) \quad 
\text{mod ${\bL}^{-\infty}$}, 
\end{eqnarray}
where $T$ is a pseudodifferential operator (PDO) in $\bL^1(\Sigma
\times I, DM)$ with principal symbol
\begin{eqnarray}\label{symbol}
\sigma_1(T)(x, \xi) = \sqrt{-h^{\mu\nu}(x)\xi_\mu\xi_\nu}
\end{eqnarray}acting `surface-wise', i.e. arises as a smooth family $\{T(t)\}_{t
\in I}$ of operators acting on the surfaces $\Sigma(t)$. 
Here $K=\nabla_\mu n^\mu$ is the extrinsic curvature
and $H(t) = -i\mslash n h^{\mu\nu} \gamma_\mu \nabla_\nu + \mslash n m$
is the Dirac Hamiltonian. 
(For the various classes of operators and symbols, we refer the reader 
to the appendix.) 
Solutions $T$ to Eq.~\eqref{factorisation}
can be found by inserting the asymptotic expansion for the 
symbol of the sought-for operator in that equation and then determining 
the terms in this expansion iteratively. 
The iterative procedure may be stopped after a finite number $n$ of 
steps, yielding operators $T_n$ which differ from $T$ by an operator
of class $\bL^{-n}$. In the case of a RW-spacetime, these will be
isotropic. A more detailed discussion of
how this works on such a spacetime and how the operators $T$ and $T_n$ 
are defined in that special case is given below, since we
do not want to interrupt the present line of argument. 

\medskip

Now fix a $t \in [t_0, t_1]$ and define 
\begin{eqnarray*}
L_{\pm}(t) = T(t) \pm H(t),  
\end{eqnarray*}
and similarly operators $L_{n, \pm}(t)$ by taking $T_n(t)$ in this 
formula. Our definition of adiabatic states is based on the following lemma, 
valid on RW-spacetimes. 
\begin{lem}\label{centlem}
Let $n \in \mn \cup \{\infty\}$. Provided $\dot R(t) \neq 0$, 
then the operators 
$L_\pm(t)$ can be modified, smoothly in $t$, by a PDO with 
smooth kernel such that there exist
hermitian, isotropic, positive operators 
$Q(t) \in {\rm OP}^{-2}$ on $L^2(\Sigma(t), DM)$
satisfying (we omit the reference to the time $t$)
\begin{eqnarray}
L_{+}QL_{+}^* + L_{-}^*QL_{-} = \myid, 
\end{eqnarray}
In the same way, if $n$ is a natural number there exists 
$Q_n$, positive and isotropic, satisfying this equation for 
$L_{n,\pm}$.
\end{lem}
Note: If $M$ is a spacetime foliated by Cauchy surfaces, then the hermitian 
adjoint of an operator acting `surface-wise' (such as in the lemma) is 
defined w.r.t. to the inner product Eq.~\eqref{innprod} on each surface.

\medskip

Before we prove the lemma, let us give the definition of 
adiabatic states for the free Dirac field on a RW-spacetime. 
Let us set
\begin{eqnarray}
\label{bdef}
B_n = L_{n,+}Q_nL_{n,+}^*. 
\end{eqnarray}
Then $B_n \in {\rm OP}^0(\Sigma(t), DM)$ is hermitian, 
isotropic and by the lemma fulfills $0 \le B_n \le \myid$.  
\begin{defn}\label{adiabdef}
The gauge invariant, quasifree states 
$\omega_n$ defined by the operators $B_n$ on $L^2(\Sigma(t), DM)$, 
$n \in \mn \cup \{\infty\}$ 
are called {\bf adiabatic states of order $n$ at time $t$}. 
\end{defn}
It is straightforward from the definitions that the
twopoint functions of such states are given by 
(suppressing the subscript $n$ for the moment)
\begin{eqnarray}\label{twopoint}
\mslash G^{(+)}(h, f) &=& 
\big\langle(in^\mu\nabla_\mu + T(t))\pgator \bar h, 
Q(t)(in^\mu\nabla_\mu + T(t)) 
\pgator f \big
\rangle_t,\nonumber\\
\mslash G^{(-)}(h, f) &=& 
\big\langle(in^\mu\nabla_\mu - T(t)^*)\pgator \bar h, Q(t)
(in^\mu\nabla_\mu - T(t)^*) 
\pgator f \big \rangle_t. 
\end{eqnarray}

\medskip
\noindent
{\bf Construction of the operators $T$ and $T_n$ in the spin-1/2 case in 
RW-spacetimes:} 
We have, 
\begin{eqnarray*}
P &=& -(i\dirac + m)(i\dirac - m)\\  
&=& -(in^\mu\nabla_\mu + iK + H)(in^\mu\nabla_\mu - H),
\end{eqnarray*} 
where we have used that $in^\mu\nabla_\mu - H = 
\mslash n(i\dirac - m)$ and the identities 
$$n^\mu\nabla_\mu \mslash n = 0,\quad 
h^{\mu\nu}\gamma_\mu\nabla_\nu\mslash n = K, $$
which follow easily 
from $n^\mu\nabla_\mu n^\nu = 0$,  
$\nabla_\mu n_\nu - \nabla_\nu n_\mu = 0$
and $\nabla_\mu \gamma_\nu = 0$. Therefore 
Eq.~\eqref{factorisation} is equivalent to the operator 
equation
\begin{eqnarray}\label{factorisationII}
H^2 + iKH + [in^\mu\nabla_\mu, H] = T^2 + iKT + [in^\mu\nabla_\mu, T] 
\quad \text{mod ${\rm OP}^{-\infty}$}  
\end{eqnarray}
for all $t \in I$, where $T$ is required to have a principal symbol
$\sqrt{-h^{\mu\nu}\xi_\mu \xi_\nu}$.  
While the above said holds for any globally hyperbolic spacetime with 
a Gau\ss ian foliation, we now specialise the discussion to RW-spacetimes, 
where one of course takes the obvious foliation by the homogeneous
surfaces $\Sigma^\kappa$. We 
consider symbols $I \times \mr \owns (t, k) \mapsto b_{ks}(t)$, 
taking values in the complex 2 by 2 matrices and carrying an 
additional helicity index $s = \pm$. We will often omit 
reference to the matrix resp. helicity indices and simply write 
$b \in \bS^n$ when we mean a matrix valued symbol of
class ${\rm S}^n(I, \mr) \otimes M_2(\mc)$. 
Any such symbol $b$ defines and operator $B$ via 
Eq.~\eqref{ansatz}. Furthermore, 
if $b \in \bS^n$ then $B \in {\rm OP}^n(\Sigma \times I, DM)$.
To see this, one can argue in just the same way as in \cite{Junker}, 
where a similar statement is proven. 

We wish to write $T$ in the form Eq.~\eqref{ansatz} with some matrix
valued symbol $\tau_{ks}$ with principal part equal to $\omega_k$. 
Inserting the ansatz Eq.~\eqref{ansatz} into 
Eq.~\eqref{factorisationII}, one obtains the following 2 by 2 matrix system of 
ordinary differential equations with parameter $k$, 
\begin{eqnarray}\label{**}
i\dot\tau + \tfrac{iN\dot R}{2R} \tau + 
[\tau, d] + \tau^2 \stackrel{\text{mod} \bS^{-\infty}}{=} 
i\dot h + \tfrac{iN\dot R}{2R} h + 
[h, d] + h^2 =: r, 
\end{eqnarray}
where $h_{ks} = {\rm diag}[\omega_k, -\omega_k]$. To arrive at this 
equation, we used Eq.~\eqref{eigenf} and 
\begin{eqnarray*}
in^\mu\nabla_\mu u^{p} = 
\sum_q i\left[{\cal U}^*\left(\tfrac{\partial}{\partial t} 
{\cal U}\right)\right]^{pq}u^{q}=: 
\sum_q d^{pq} u^{q},  
\end{eqnarray*}
where 
\begin{eqnarray}\label{christof}
\label{matid}d_{ks}^{pq} = 
\left[
\begin{matrix}
 0 & -\frac{iskm\dot R}{2(m^2 R^2 + k^2)} \\
\frac{iskm\dot R}{2(m^2 R^2 + k^2)} & 0
\end{matrix}
\right].
\end{eqnarray}
The drop-off  properties in $k$ required from the the matrix 
$\tau_{ks}$ uniformly in the time interval $I$, imply that the 
initial data must be carefully adjusted, and cannot be freely
chosen.  

We try to find $\tau$ as
an asymptotic expansion of its symbol (e.g. in a certain 
sense `in powers of $k$'). First note, that since
Eq.~\eqref{factorisation} should hold only up to the addition of a
arbitrary operator of class $\bL^{-\infty}(\Sigma \times I, DM)$, $\tau$ need
only be defined up to $\bS^{-\infty}$. It is therefore enough to 
find an asymptotic expansion for $\tau$, 
\begin{eqnarray*}\label{asymptotic}
\tau \sim \sum_j \vartheta_j, 
\end{eqnarray*}
where $\vartheta_j \in \bS^{1-j}$ and matrix valued. 
In order to get an operator with the right principal symbol we set
$\vartheta_{0, sk}(t) = \omega_k(t)$ for all $t\in I$. 
We then define $\vartheta_j$ successively in such a way
that the $n$'th partial sum 
$\tau_n=\sum_j^n \vartheta_j$ in Eq.~\eqref{asymptotic} solves 
Eq.~\eqref{**} modulo $\bS^{1-n}$. This can be achieved by
putting
\begin{eqnarray}\label{recursion}
\vartheta_{n+1} = \frac{-1}{2\vartheta_0}\left( i\dot\tau_{n} + 
[\tau_{n}, d] + \tau^2_{n} + 
\tfrac{iN\dot R}{2R} \tau_n - r \right)  \quad 
\in \bS^{-n}(I, \mr).
\end{eqnarray}
Following this procedure we can calculate terms of arbitrary high 
order in the asymptotic expansion for $\tau$. 
Any symbol with this expansion will give rise to an 
operator $T$ factorising the spinorial KG-operator 
modulo ${\rm OP}^{-\infty}$. 
Moreover, the partial sum $\tau_n$ obtained after 
$n$ iterations will give rise to an isotropic operator $T_n$, which will
solve Eq.~\eqref{factorisation} up to $\bL^{-n}$.  
$T$ as well as $T_n$ have the same principal symbol as
$|H|$, i.e. Eq.~\eqref{symbol} holds true. 

\medskip

Clearly, if the existence of operators 
$Q_n$ as in the lemma is known for a general globally hyperbolic spacetime, 
then one can still define operators $B_n$ by Eq.~\eqref{bdef}, and these in  
turn give quasifree, gauge invariant states $\omega_n$. 

\begin{prop}\label{chargecon}
Suppose $(M, g)$ is a spacetime with a compact 
Cauchy surface $\Sigma$, and suppose there exist operators $Q_n$ ($n \in 
\mn \cup \{\infty\}$) as in 
the above lemma. Then there are operators $0 \le \tilde B_n \le \myid$, 
differing from $B_n$ only by an operator of class ${\rm OP}^{-n-1}$,
such that $C\tilde B_n C^{-1} = \myid - \tilde B_n$ for all $n$. 
\end{prop} 
\begin{remark}
It follows from the definition of the charge conjugation automorphism
that the states $\tilde \omega_{n}$ corresponding to $\tilde B_n$ are 
charge invariant.
\end{remark}
\begin{proof}
We treat the case of infinite order first and suppress the subscript
$n$. We also introduce the notation 
$A^c = CAC^{-1}$ for operators $A$ acting
on spinors. We need to show that $B$ can be modified modulo 
$C^\infty$ to an operator $\tilde B$ such that $\tilde B^c = \myid -
\tilde B$ and $0\le \tilde B \le \myid$. 
Taking the transpose of Eq.~\eqref{factorisation} we obtain
\begin{eqnarray*}
P^t = -(-in^\mu\nabla_\mu - iK - T^t)(-in^\mu\nabla_\mu + T^t) 
\quad \text{mod $\bL^{-\infty}$.} 
\end{eqnarray*}
It is not difficult to see that $\overline{Pf} = P^t\bar f$ and 
$T^t \bar f = \overline{T^* f}$.
From this it follows that
\begin{eqnarray}\label{factorisation*}
P = -(in^\mu\nabla_\mu + iK - T^*)(in^\mu\nabla_\mu + T^*) 
\quad \text{mod ${\rm OP}^{-\infty}$}
\end{eqnarray}
But multiplying Eq.~\eqref{factorisation} with $C$ from both 
sides and using that 
$(in^\mu\nabla_\mu)^c = -in^\mu\nabla_\mu$, we also 
have that
\begin{eqnarray*}
P = -(in^\mu\nabla_\mu + iK - T^c)(in^\mu\nabla_\mu + T^c) 
\quad \text{mod $\bL^{-\infty}$.}
\end{eqnarray*}
Therefore, since the principal symbols of 
$T^*$ and $T^c$ are equal and since the factorisation is 
unique modulo ${\rm OP}^{-\infty}$ once this information 
is known, we have shown that
\begin{eqnarray*}
T^* = T^c \quad \text{mod ${\rm OP}^{-\infty}$.}
\end{eqnarray*}
We redefine $T$ by  
$\frac{1}{2}(T + T^{*c})$ and $Q$ by  $\tfrac{1}{2}(Q + Q^c)$ and
$L_\pm$ by inserting the modified definition of $T$. 
These redefined operators will then satisfy 
(remembering that that $H^c= -H$ and using that 
$A^{c*} = A^{*c}$ for any operator $A$ acting on spinor fields 
over $\Sigma(t)$\footnote{
Here one must use that the foliation is Gau\ss ian. }.)
\begin{eqnarray*}
Q^c = Q, \quad  Q \ge 0, \quad L_\pm^c = L_\mp^*
\end{eqnarray*} 
and 
\begin{eqnarray*}
L_+Q L_+^* + L_-^* Q L_- = \myid + A, 
\end{eqnarray*} 
where $A \in {\rm OP}^{-\infty}$. 
Since the left hand side of this equation is invariant under 
charge and hermitian conjugation and positive, we find 
$A^c = A$, $A^* = A$ and $\myid + A \ge 0$. 
Since $\Sigma(t)$ is compact, $A$ is a compact operator and 
the projectors on all nonzero eigenspaces have a smooth kernel. 
Let $F$ be the projector on the 
(finite dimensional) kernel of $\myid + A$. 
Then $R = \myid + A + F$ is strictly positive, 
$[F, R] = 0$, $FR = F$,  $F^c = F$, $R^c = R$. Let us write
\begin{eqnarray*}
\tilde L_\pm = R^{-1/2}L_\pm R^{-1/2}, \quad 
\tilde Q = R^{1/2} Q R^{1/2}. 
\end{eqnarray*} 
Then $\tilde L_\pm^c = \tilde L_\mp^{*}$, 
$\tilde Q \ge 0$, and the operator $\tilde B$ defined 
by 
\begin{eqnarray*}
\tilde B = \tilde L_+ \tilde Q \tilde L_+^{*} + 
\tfrac{1}{2}F
\end{eqnarray*}
satisfies  $B = \tilde B$ modulo smooth, 
$\tilde B^c = \myid - \tilde B$ and 
$0 \le \tilde B \le \myid$, providing us thus with a modified operator 
with the desired properties. For arbitrary 
$n \ge 1$, one proceeds in a similar way, this time 
using that $T_n^* = T_n^c$ modulo ${\rm OP}^{-n-1}$.
\end{proof}

\medskip
\noindent
Proof of lemma~\ref{centlem}:
\begin{proof}
The argument establishing the existence of $Q_n$ as in the lemma 
is the same for all $n \ge 1$, therefore, to lighten the 
notation we will only treat the case 
$n = \infty$ and drop the reference to $n$. 
We set $\ell_{\pm} = \tau \pm h$. The $\ell_{\pm}$ are then 
related to the PDO's $L_\pm$ in the statement of the lemma by 
Eq.~\eqref{ansatz}. One finds after the first iteration, 
\begin{eqnarray*}
\ell_{+,ks} = 2\left[
\begin{matrix}
\omega & 0 \\
0  &-\frac{i\partial_t(R^{N/2}\omega_k)}{R^{N/2}\omega_k}
\end{matrix}
\right], 
\quad 
\ell_{-,ks} = 2\left[
\begin{matrix}
0 & 0 \\
0  & \omega_k - \frac{i\partial_t(R^{N/2}\omega_k)}{R^{N/2}\omega_k}
\end{matrix} 
\right] 
\end{eqnarray*}
modulo $\bS^{-1}$. Further iterations change $\tau$ only by 
symbols of order 
less or equal $-1$ and will therefore not affect the above form of 
$\ell_\pm$. Only the above form of the $\ell_\pm$
is used to argue the existence of $Q$, therefore the adiabatic 
order $n$ is not important for 
our argument, as long as $n \ge 1$ . The proof of the lemma then 
amounts to show that $\tau$ can be modified by a symbol of 
class $\bS^{-\infty}$ such that one can find a 
$q \in {\rm S}^{-2}$(or rather one for each helicity), taking 
values in the complex 2 by 2 matrices, such that 
($^*$ denotes the hermitian adjoint of a matrix and $\myid$ 
the identity matrix)
\begin{eqnarray}\label{comp}
\ell_{+,ks} q_{ks} \ell_{+,ks}^* + \ell_{-,ks}^* q_{ks} \ell_{-,ks} 
= \myid, \quad q_{ks} = q^*_{ks}, \quad q_{ks} \ge 0 \quad \forall k \ge 0, s = \pm.
\end{eqnarray} 
The PDO $Q$ corresponding to $q$ by Eq.~\eqref{ansatz} 
then obviously fulfills the claim of the lemma 
(note that the integral/sum in Eq.~\eqref{ansatz} is over 
positive $k$ only). Taking the matrix adjoint of Eq.~\eqref{comp}, 
one observes that $q$ can be taken to be hermitian. 
One may regard Eq.~\eqref{comp} as a linear equation for $q$ at 
each value of $k$ and the helicity index $s$, and thus write it 
as a 4 by 4 matrix system for the matrix entries of $q$ 
\begin{eqnarray}\label{matrixe}
M_{ks} \left[
\begin{matrix}
q^{11}_{ks}\\
q^{22}_{ks}\\ 
q^{12}_{ks}\\ 
q^{21}_{ks}
\end{matrix}
\right]
= 
\left[
\begin{matrix}
1\\
1\\
0\\
0
\end{matrix}
\right], 
\end{eqnarray}
where $M_{ks}$ is a 4 by 4 matrix determined from the entries of $\ell_{\pm,ks}$.
Only using the above from of $\ell_{\pm}$ and the fact that
${\rm S}^m {\rm S}^n \subset {\rm S}^{n+m}$ one finds from 
Eq.~\eqref{comp}
\begin{eqnarray*}
M &=& D(\myid + A) \quad \text{where}\\
D_{ks} &=& 4\,{\rm diag}[\omega^2_k, \omega^2_k,
-iR^{-N/2}\partial_t(R^{N/2}\omega_k),
iR^{-N/2}\partial_t(R^{N/2}\omega_k)], \quad  
A \in \bS^{-1}.
\end{eqnarray*}
Hence, if $\dot R(t) \neq 0$, then $M_{ks}$ has a matrix inverse 
in $\bS^{-2}$ for large $k$, 
\begin{eqnarray*}
M^{-1}_{ks} = D^{-1}_{ks} \sum_{m=0}^\infty A^m_{ks}.
\end{eqnarray*}
Eq.~\eqref{matrixe} may therefore be inverted for large 
$k$ and gives us a solution $q$ to Eq.~\eqref{comp}. 
It follows directly from the above form of $M^{-1}$ that $q_{ks}$ has
${\rm diag}[(2\omega_k)^{-2},(2\omega_k)^{-2}]$ as a principal 
symbol, therefore $q_{ks}$ is positive 
definite for large $k$. We have thus constructed a solution to 
Eq.~\eqref{comp} if $k$ is greater than some $k_0 \ge 0$. 
To find such a $q$ also for $0 \le k \le k_0$, we may redefine
\begin{eqnarray*}
\tau(t, k) = 
\begin{cases}
\tau(t, k_0) & \text{for $0 \le k \le k_0$},\\
\tau(t, k) &   \text{for $\quad k \ge k_0$,}
\end{cases}
\end{eqnarray*} 
and arbitrarily for $k \le 0$, since $T$ by 
definition does not depend on $\tau(t, k)$ for $k \le 0$.
By what we have already shown, such a $\tau$ trivially allows 
for a hermitian, positive solution of Eq.~\eqref{comp}, but it 
is not yet a symbol (because its dependence on $k$ is not 
smooth). We might however change the above definition
of $\tau$ in an arbitrary small neighbourhood of $k_0$ to make 
it smooth (and hence a symbol), without making the corresponding
matrix $M$ singular. As we mentioned earlier,  
the resulting matrix $q$ is automatically hermitian. By easy 
arguments based on the continuity of the construction, it will 
also remain positive, if that change is made
arbitrarily small.  
\end{proof}

\section{Properties of adiabatic states}
\begin{prop}\label{propo2}
The states $\omega_n$ are locally quasiequivalent to a Hadmard
state if $n \ge N$. Furthermore, the difference between the
twopoint functions $\mslash G_n$ of $\omega_n$ and those of 
a Hadamard state is given by a $C^{n-N+1}$ kernel.
\end{prop}
\begin{proof}
According to Thm.~\ref{adahad}, an adiabatic state of infinite
order (defined, as described above, by a symbol $b \in 
\bS^0(I, \mr)$) is Hadamard, so it is sufficient to
show that adiabatic states of order $n$ (described
by a symbol $b \in \bS^0(I, \mr)$) are locally quasiequivalent
to such a state.  Now $b - b_n$
is by definition a symbol of order $-(n+1)$, so in particular,
\begin{eqnarray*}
\left\|b_{ks} - b_{n \, ks} \right\| \le 
c|1 + k|^{-n-1} \quad\text{for all 
$k \ge 0$.}
\end{eqnarray*}
The criterion on local quasiequivalence, Thm.~\ref{qeq} then immediately
proves that the states are locally quasiequivalent if $n \ge N$. 

Let $A_n\in \bL^{-n-1}(\Sigma(t), DM)$ be the operator associated to the 
symbol $a_n = b - b_n$ via Eq.~\eqref{ansatz} at some $t \in I$. 
By standard theorems, e.g. in~\cite[II, Prop. 2.7]{tay}, the associated
kernel on $\Sigma(t) \times \Sigma(t)$ is in  $C^{n-N+1}$ 
for $n \ge N-1$. The difference of the twopoint function
of an adiabatic state of infinite order and one of order 
$n$ is
\begin{eqnarray*}
\mslash G^{(\pm)} (f, h) - \mslash G^{(\pm)}_n (f, h) 
= \pm \langle (\pgator \bar h) \restriction \Sigma(t) 
, A_n (\pgator f) \restriction \Sigma(t) \rangle_t. 
\end{eqnarray*}
Now the causal propagator $\pgator$  
propagates $m$ times differentiable initial data to 
$m$ times differentiable solutions. 
Therefore the above difference must $(n - N + 1)$ times 
differentiable in $M \times M$.  
\end{proof}

\begin{thm}\label{adahad}
The adiabatic states are Hadamard in the sense of
Def.~\ref{Hadamardcond}. 
\end{thm}
\begin{proof}
In order to prove that $\mslash G^{(\pm)}$ has the wave front 
set described in 
Def.~\ref{Hadamardcond}, we shall employ the following result due to 
W.~Junker~\cite[Thm 3.12]{Junker}.
This result has originally been obtained for scalar fields, but a 
careful analysis of the proof shows that it can be adapted to the 
spinor case. We present here 
a modified version which is tailored to our situation. $E$ is the
causal propagator for the spinorial Klein-Gordon operator $P$.   
\begin{thm}
Let $Q(t)$ be an elliptic PDO on $\Sigma(t)$. Let $I$ be an interval
containing $t$ and $A_\pm\in\bL(\Sigma \times I, DM)$ such that there 
exist PDO's $R_\pm \in \bL(\Sigma \times I, DM)$ which have the 
property $R_\pm(in^\mu\nabla_\mu + A_\pm) = P$ modulo smooth and 
\begin{eqnarray*}
{\cal Q}_{R_\pm} \subset \{(x, \xi) \in T^*M \backslash \{0\} \mid \xi \triangleright 
(\triangleleft) \,0\},  
\end{eqnarray*}
where ${\cal Q}_{R_\pm}$ is defined in Eq.~\eqref{qdef}. 
Then the spinorial bidistributions 
\begin{eqnarray*}
\mslash \Lambda^{(\pm)}(h, f) 
= \langle (in^\mu\nabla_\mu + A_\pm(t)) E \bar h
, Q(t) (in^\mu\nabla_\mu + A_\pm(t)) E f
\rangle_t
\end{eqnarray*}
have wave front set $\WF'(\mslash\Lambda^{(\pm)}) \subset C^{(\pm)}$. 
\end{thm}
We apply the lemma to $A_- = -T$ and $A_+ = T^*$ 
and $Q$ as in the definition of the twopoint functions, 
Eq.~\eqref{twopoint}. Then Eq.~\eqref{factorisation}, 
and Eq.~\eqref{factorisationII} provide us with operators 
\begin{eqnarray*}
R_- = -(in^\mu\nabla_\mu + iK + T), \quad 
R_+ =-(in^\mu\nabla_\mu + iK - T^*)
\end{eqnarray*}  
as in the statement of the above theorem.
Clearly, since 
$\sigma_1(T) = \sigma_1(T^*) = \sqrt{-h^{\mu\nu}\xi_\mu\xi_\nu}$, 
\begin{eqnarray*}
{\cal Q}_{R_\pm} = 
\{ (x,\xi) \mid n^\mu\xi_\mu = 
\pm \sqrt{-h^{\mu\nu}\xi_\mu\xi_\nu} \}
= \{(x, \xi) \mid \xi \triangleright (\triangleleft) 0\}
\end{eqnarray*} 
Noting that $(-i\dirac - m)E
= \mslash S$ and using the fact that the wave front set cannot become
larger upon acting with a PDO on a distribution, we can apply 
Junker's theorem to $G^{(\pm)}$ (given by Eq.~\eqref{twopoint}) 
and obtain $\WF(\mslash G^{(\pm)}) \subset C^{(\pm)}$. 
It remains to show equality in the above
inclusions. The anticommutation relations imply that 
$\mslash G^{(+)} + \mslash G^{(-)} = i\pgator$. If the 
causal propagator $\mslash S$
had wave front set $W = \{ (x_1, \xi_1, x_2, \xi_2) \mid 
(x_1, \xi_1) \sim (x_2, \xi_2)\} = C^{(+)} \cup C^{(-)}$ 
(and this will indeed be shown) then
\begin{eqnarray*}
W = \WF'(\pgator) \subset \WF'(\mslash G^{(+)}) \cup 
\WF'(\mslash G^{(-)})
\subset C^{(+)} \cup C^{(-)} = W,
\end{eqnarray*}
thus in fact equality would hold in the above inclusions. We have to
show that $\mslash S$ has indeed wave front set $W$. From $\mslash S^c
= \mslash S$ and the {\it antilinearity} of the charge conjugation 
it follows that
\begin{eqnarray}\label{chargeinv}
\WF'(\mslash S) = -\WF'(\mslash S).
\end{eqnarray}
One has (see~\cite{Dim}), 
$\mslash S \restriction\Sigma\times\Sigma = i\mslash n \myid$
where $\myid$ means
the identity on the Cauchy surface. From \cite[Thm. 2.22]{Junker}, 
one knows that
\begin{eqnarray*}
\WF'(\mslash S{\restriction\Sigma\times\Sigma}) &\subset& d\phi^t \circ
\WF'(\mslash S) := \{
  (x_1, d\phi_{x_1}^t(\xi_1), 
   x_2, d\phi_{x_2}^t(\xi_2)) \mid\\
&& (\phi(x_1), \xi_1, 
   \phi(x_2), \xi_2) \in \WF'(\mslash S), \quad x_1, x_2 \in \Sigma \}
\end{eqnarray*}
where $\phi: \Sigma \to M$ is the embedding map. Now assume that 
$(x_1, \xi_1, x_2, \xi_2)$ is in $W$ but not in the wave front set of 
$\mslash S$. By the propagation of singularities, we can assume that
there is an element $(x, \xi, x, \xi)$, $x \in \Sigma$, 
which is in $W$ but not in $\WF'(\mslash S)$. By Eq.~\eqref{chargeinv}, also 
$(x, -\xi, x, -\xi) \notin \WF'(\mslash S)$. 
Since $\xi$ must be a nonzero null covector, it is 
impossible that the nonzero 
element $(x, d\phi_{x}^t (\xi), x, d\phi_{x}^t (\xi))$ 
is in $d\phi^t \circ \WF'(\mslash S) \supset \WF'(i\mslash n \myid)$. 
But the latter set is actually equal to 
\begin{eqnarray*}
\WF'(\delta^{(N)}) = 
\{(x, \xi, x, \xi) \mid \xi \in T^*\Sigma
\backslash \{0\}\}, 
\end{eqnarray*}
and so must contain any element of that form, a contradiction. 
\end{proof}

\section{Comments}

\begin{enumerate}
\item
At adiabatic order zero we get $B_0(t) = \tfrac{1}{2} |H(t)|^{-1}(H(t) + |H(t)|)$. 
i.e. $B_0(t)$ projects on the instantaneous positive frequency solutions at time $t$. 
Clearly, if $R$ is constant in a neighbourhood of $t$, then there are no further 
corrections to this operator at higher adiabatic orders, and $B_0(t)$ defines
a pure quasifree Hadamard state. If $R$ is not constant near $t$, 
then the operators $B_n(t)$ give states which are not Hadamard in general 
but {\it only reproduce the highest order singularities of the Hadamard form}, 
in the sense that\footnote{
Note that the next term in this series would be 
$\sigma^{n-N+3} \log \sigma$, which is $(n-N+2)$ times 
differentiable.} (modulo $C^{n-N+1}$),  
\begin{multline*}
\mslash G^{(\pm)}_{n}(x_1, x_2) 
=\\
\pm(i\dirac + m)
\left[
\sum_{j = 1}^{(N-1)/2} U_j(x_1,
x_2)\sigma_{\pm\epsilon}^{-j} + \sum_{j = 0}^{n-N+2} 
V_j(x_1, x_2) \sigma^j \log \sigma_{\pm\epsilon} 
\right].
\end{multline*} 
Hence, for $n \ge N$, adiabatic states 
will allow for a point-splitting renormalisation of the stress-energy 
tensor $T_{\mu\nu}$ as described e.g. in \cite{wald} (the 
difference to a Hadamard state must be at least in $C^1$, 
because the stress tensor contains 1 derivative), but such of lower 
order will not in general. 
{\it In other words, we see that any adiabatic vacuum state which 
allows for a point splitting renormalisation of the stress-energy 
tensor will be locally quasiequivalent to a Hadamard state.}

\item
In \cite{no}, the authors observe that the expected stress tensor
of a Dirac field diverges for the 
`energy-minimising states' proposed in that work. 
This is explained by our analysis, since their 
states are simply the adiabatic states of order zero 
just discussed in diguise. 

\item
Following the strategy of
L\"uders and Roberts~\cite{lr} for the scalar Klein-Gordon field, \cite{wellm}
proposes another definition of adiabatic states for the Dirac field. 
We do have some doubts as to whether their definition really
yields a positive state, moreover the r\^ole of positive and negative
frequency modes remains obscure in \cite{wellm}. It is therefore 
difficult to see how their definition relates to ours. In view of the
analysis carried out in this paper, we do not believe that the 
states proposed in \cite{wellm} are of Hadamard type, as suggested by the author.

\item
It is possible to construct a Hadamard state in a general globally 
hyperbolic spacetimes along the same lines as in the previous
section if one can construct a hermitian, positive
PDO $Q$ such that 
\begin{eqnarray*}
L_+QL_+^* + L_-^*QL_- = \myid.
\end{eqnarray*}
However, unlike in the simple case of a 
RW-spacetime, the construction of such a $Q$ seems to be 
harder. We are currently working on this problem. 
\end{enumerate}

\section{Appendix}

\subsection{Spinors on flat space and representation theory}

In this appendix we find the (generalised) eigenfunctions of the spatial
Dirac operator $\widetilde \dirac$ on $\mr^N$. To this end, we first write
this operator in polar coordinates, 
\begin{eqnarray*}
\widetilde\dirac \psi = \left( \partial_\theta + \frac{N-1}{2\theta} \right)
\gamma^N \psi + \frac{1}{\theta} \widetilde\dirac_{N-1}\psi, 
\end{eqnarray*}
where $\widetilde\dirac_{N-1}$ is the Dirac operator on $\ms^{N-1}$. 
The eigenfunctions of this operator \cite{Trautman}, 
\begin{eqnarray*}
\widetilde\dirac_{N-1} \xi^{(\pm)}_{lm} = \pm i(l + (N-1)/2)\xi^{(\pm)}_{lm}, 
\quad l = 0, 1, \dots, \,\, m = 0, 1, \dots, d_l,   
\end{eqnarray*}
may be used to find the spectral decomposition of Dirac operator on
$\mr^N$. We first set
\begin{eqnarray*}
\hat\xi_{lm}^{(\pm)} = \frac{1}{\sqrt{2}}\left(\xi^{(-)}_{lm} \pm 
i\gamma^N \xi^{(+)}_{lm}\right).   
\end{eqnarray*}
In order to find the eigenfunctions of $\widetilde\dirac$ we insert the
ansatz
\begin{eqnarray}\label{Chi}
\chi_{klm\pm}(\theta, \Omega) = c(kl)\left(a_{kl}(\theta) \hat\xi^{(+)}_{lm}
(\Omega) \pm ib_{kl}(\theta) \hat\xi^{(-)}_{lm}(\Omega)
\right), \quad \Omega \in \ms^{N-1}
\end{eqnarray}
into 
\begin{eqnarray*}
\widetilde\dirac\widetilde\dirac \chi_{klms} = -k^2\chi_{klms}.  
\end{eqnarray*}
This leads to the differential equation  
\begin{eqnarray*}
\left[\partial_\theta^2 + \frac{N-1}{\theta}\partial_\theta + 
\frac{l(l + N - 2)}{\theta^2} + k^2\right] a_{kl}(\theta) = 0, 
\end{eqnarray*}
for $a_{kl}$ (and similarly $b_{kl}$). The unique regular solutions to 
these equations are given by Bessel functions, 
\begin{eqnarray}\label{ab}
a_{kl}(\theta) &=& \theta^{-(N-2)/2} J_{l+(N-2)/2}(k\theta)\nonumber\\
b_{kl}(\theta) &=& \theta^{-(N-2)/2} J_{l+N/2}(k\theta). 
\end{eqnarray}
The normalisation factor in Eq.~\eqref{Chi} is determined from the condition 
\begin{eqnarray*}
\langle \chi_{klms}, \chi_{k'l'm's'} 
\rangle = 
\delta(k - k')\delta_{ll'}\delta_{mm'}\delta_{ss'}, 
\end{eqnarray*} 
one finds $c(kl)=\sqrt{k/2}$.
In this work we also need the spectral function (Plancherel measure)
defined by
\begin{eqnarray*}
P_N(k) = \sum_{lm}  \chi_{klms}(0)^\dagger \chi^{(s)}_{klms}(0).   
\end{eqnarray*}
From the expression Eq. \eqref{ab} and behaviour of Bessel functions at $\theta = 0$
it is seen that only the term with $l = 0$ will contribute, 
leading to the result
\begin{eqnarray*}
P_N(k) = \frac{k^{N-1}}{2^{N/2}\,{\rm vol}(\ms^{N-1})\Gamma(N/2)^2}. 
\end{eqnarray*}

\subsection{Notions and results from microlocal analysis}

For convenience we mention some results and definitions 
from the theory of distributions and the theory of pseudodifferential operators
(PDO's). 
If not indicated otherwise, these may be found in standard textbooks, for example
\cite{tay, hodu}. PDO's generalise ordinary differential operators in 
the sense that they give meaning to fractional
powers of derivatives. They are defined in terms of so-called symbols. 
We shall not give the most general definition of a symbol here, since
only a certain class of symbols is important for this work. 

\begin{defn}
Let $\cO$ be a subset of $\mr^n$ and $m$ be a real number. Then a symbol 
of order $m$ is a function $a \in C^\infty(\cO, \mr^n)$ such that 
for every compact
subset $K$ of $\cO$ the following estimate holds 
\begin{eqnarray*}
\left|D^\alpha_x D^\beta_\xi a(x, \xi) \right|\le C_{\alpha, \beta, K}
(1 + |\xi|)^{m - |\beta|}
\end{eqnarray*}
for all multiindices $\alpha, \beta$. $D^\alpha$ is $i^{|\alpha|}
\partial_1^{\alpha_1} \dots \partial_n^{\alpha_n}$.
The set of all such symbols is denoted by $S^m(\cO, \mr^n)$ and  
one also writes $\bS^{-\infty} = \bigcap_m \bS^m$. 
\end{defn}
There is the notion of the asymptotic expansion of a 
symbol which is an important tool for constructing PDO's. 
Suppose $a_j \in \bS^{m_j}(\cO, \mr^n)$ for $j = 0, 1, 2, \dots$ 
with $m_j$ monotonously decreasing to minus infinity. Then there exists
$a \in \bS^{m_0}(\cO, \mr^n)$ such that for all $N$ 
\begin{eqnarray*}
a - \sum_{j = 0}^N a_j \in \bS^{m_N}(\cO, \mr^n)
\end{eqnarray*}
and $a$ is defined modulo $\bS^{-\infty}$. One writes $a \sim \sum_j a_j$.
If $a \in \bS^{m}(\cO, \mr^n)$ then the operator
\begin{eqnarray*}
Au(x) = \int e^{ix\xi} a(x, \xi) \hat u(\xi) \frac{d^n\xi}{(2\pi)^n}
\end{eqnarray*} 
is said to belong to $\bL^m(\cO)$, the PDO's of 
order $m$. $A$ is a continuous linear operator from $\cD(\cO)$ to
$C^\infty(\mr^n)$. By the Schwartz kernel theorem it is thus given
by a distribution kernel $K_A \in \cD'(\cO \times \cO)$. $K_A$ is 
smooth off the diagonal in $\cO \times \cO$ and smooth everywhere 
in $\cO \times \cO$ if $A \in \bL^{-\infty}(\cO)$. Hence the asymptotic 
expansion of a symbol uniquely determines a PDO modulo smoothing operators. 
The above statement carries over to matrix valued symbols without 
major changes. A principal symbol $\sigma_m(A)$ of $A \in \bL^m(\cO)$
is a representer of its symbol in $\bS^m(\cO)/\bS^{m-1}(\cO)$. 
It can be chosen such that it transforms contravariantly under a 
change of coordinates (giving thus a well-defined function on the
cotangent bundle) and it behaves multiplicatively under multiplication of
two PDO's. On a manifold $M$ (or more generally on a vector-bundle $E$) 
PDO's are defined to be the continuous operators on 
$\cD(M, E)$ which have the above properties in each coordinte patch. 

We come to the definition of the polarisation set of a vector-valued 
distribution $u = (u^1, \cdots, u^k) \in \cD'(\cO)^k$, $\cO$ an
open subset of $\mr^n$. For details of the definition and the subsequent
results see~\cite{Denck}.  
\begin{defn}
The ``polarisation set'' $\Pol(u)$ of a vector-valued distribution $u$ is defined as
\begin{eqnarray*}
\Pol(u) = \bigcap_{A \in \bL^0, \,\, Au \in C^\infty} \cN_A,  
\end{eqnarray*}
where 
\begin{eqnarray*}
\cN_A = \{(x, \xi, w) \in T^*\cO \times \mc^k \mid
\sigma_0(A)(x,\xi)w = 0 \}. 
\end{eqnarray*}
\end{defn}
From the transformation properties of the principal symbol it is clear
that the definition can be carried over to the case of distributions
with values in a vector-bundle $E$. $\Pol(u)$ is then seen to be a
linear subset of $\pi^* E$, $\pi: T^*M \rightarrow M$ being the canonical 
projection in the fibres of the cotangent bundle. The ``wave
front set'' $\WF(u)$ of a distribution is obtained by taking all points
$(x, \xi) \in T^*M$ such that the fibre over this point in
$\Pol(u)$ is nontrivial. The microlocal properties of the bidistributions
considered in this work are more conveniently described in terms
of their primed polarisation set, $\Pol'$, which is obtained from the usual
one by reversing the sign of the covectors in the second slot (the 
primed wave front set is defined similarly).

There is an important theorem on the polarisation set of distributions 
$u$ satisfying $Pu \in C^\infty$ for differential operators $P$ of real 
principal type, which goes under the name `propagation of singularities'
\cite{Denck, hodu}. Such operators are defined as follows (in the following 
we discuss the simple case where $P$ acts in the bundle $E = M 
\times \mc^k$, but all results can be generalised in the obvious way
to nontrivial vector bundles):
\begin{defn}\label{realprincipal}
A $k \times k$ system $P$ of differential operators on 
a manifold $M$ with principal symbol $p_0(x, \xi)$ is said to be
of real principal type at $(y, \eta)$ if there exists a 
$k \times k$ symbol $\tilde p_0(x, \xi)$ such that 
\begin{eqnarray*}
\tilde p_0(x, \xi) p_0(x, \xi) = q(x, \xi)1_k
\end{eqnarray*} 
in a neighbourhood of $(y, \eta)$, where $q(x, \xi)$ is scalar and
of scalar real principal type, i.e., $\partial_\xi q(x, \xi) \neq 0$
for all $\xi \neq 0$. 
\end{defn}
One sets
\begin{eqnarray}\label{qdef}
{\cal Q}_P = \{ (x, \xi) \mid
{\rm det}p_0(x, \xi) = 0 \}. 
\end{eqnarray}
If $f$ is a $C^{\infty}$ function on ${\cal Q}_P$ with values in
$\mc^k$, then one defines
\begin{eqnarray} \label{dencon}
D_Pf = X_qf + \frac{1}{2}\{\tilde p_0, p_0\} f + i\tilde p_0 p^s f, 
\end{eqnarray}
$X_q$ being the Hamiltonian vector field of $q$, 
\begin{eqnarray*}
X_q &=& \partial_x q \, \partial_\xi -
        \partial_\xi q \, \partial_x, \quad
\{ \tilde p_0, p_0 \} = \partial_\xi \tilde p_0 \, \partial_x p_0 -
                        \partial_x \tilde p_0 \, \partial_\xi p_0,\\
p^s &=& p_1 + 
\frac{1}{2}\partial_\xi D_x p_0, \quad
\sigma(P) \sim p_0 + p_1 + p_2 + \dots.
\end{eqnarray*} 
One can prove that $D_P$ is a partial connection along the 
Hamiltonian vector field  restricted to ${\cal Q}_P$. 
Since there is some arbitrariness in the choice of the symbol $\tilde p$,
the partial connection is not uniquely defined. One can however 
prove that the remaining arbitrariness is irrelevant in what follows. 
\begin{defn}
A Hamilton orbit of a system $P$ of real principal type is 
a line bundle $\cL_P \subset \cN_P \restriction c$, ($c$ is an
integral curve of the Hamiltonian field on ${\cal Q}_P$, 
$\dot c(t) = X_q(c(t))$) which is spanned by a sections $f$ satisfying $D_Pf = 0$, 
i.e. $\cL_P$ is parallel with respect to the partial connection. 
\end{defn}
\begin{thm}\label{propsing}
Let $P$ be of real principal type and $u$ a vector-valued 
distribution. Suppose $(x, \xi) \notin \WF(Pu)$. Then, over a 
neighbourhood of $(x, \xi)$ in ${\cal Q}_P$, $\WF_{pol}(u)$ is a
union of Hamilton orbits of $P$.
\end{thm}

{\bf Aknowledgements:} I would like to thank K. Fredenhagen, C.J. Fewster,
B.S. Kay and especially M. Radzikowski and W. Junker for helpful
discussions and comments. I am also grateful to K. Kratzert for pointing
out his work to me and for noticing an error in the proof of 
Thm.~\ref{poltheorem} in an earlier version of this paper.

\end{document}